\documentclass[orivec]{llncs-compact}

\usepackage{latexsym,amssymb,amsmath}
\usepackage{url}
\usepackage[defblank]{paralist}
\usepackage{tikz,subfigure}
\usetikzlibrary{arrows,shapes,shapes.multipart,snakes,automata,backgrounds,petri,positioning,shadows,matrix,decorations.pathmorphing,fit,positioning,calc,backgrounds}
\usepackage{todonotes}
\usepackage{xspace}
\usepackage{pifont}
\usepackage{fixltx2e}
\usepackage{listings}
\usepackage{wrapfig}

\newcommand{\dbnet}{DB-net\xspace}
\newcommand{\dbnets}{DB-nets\xspace}

\newcommand{\dbn}{\mathcal{B}}
\newcommand{\nucpn}{\ensuremath{\nu}-CPN\xspace}
\newcommand{\nucpns}{\ensuremath{\nu}-CPNs\xspace}

\newcommand{\set}[1]{\{#1\}}
\newcommand{\tup}[1]{\langle#1\rangle}
\newcommand{\mult}[1]{#1^\oplus}

\newcommand{\true}{\mathsf{true}}
\newcommand{\false}{\mathsf{false}}

\newcommand{\funsym}[1]{\mathtt{#1}}
\newcommand{\setsym}[1]{\mathit{#1}}
\newcommand{\restr}[2]{{
  \left.\kern-\nulldelimiterspace 
  #1 
  \vphantom{\big|} 
  \right|_{#2} 
  }}

\newcommand{\R}{\mathcal{R}}
\newcommand{\E}{\mathcal{E}}

\newcommand{\M}{\mathcal{M}}
\newcommand{\I}{\mathcal{I}}

\newcommand{\schema}{\R}

\newcommand{\adom}[2][]{\setsym{Adom}_{#1}(#2)}

\newcommand{\free}[1]{\setsym{Free}(#1)}

\newcommand{\subst}{\theta}

\newcommand{\relname}[1]{\ensuremath{\mathit{#1}}\xspace} 
\newcommand{\cname}[1]{\ensuremath{\mathtt{#1}}\xspace} 

\newcommand{\add}{\mathbf{add}}
\newcommand{\del}{\mathbf{del}}

\newcommand{\action}{\alpha}
\newcommand{\doact}[2]{\funsym{apply}(#1,#2)}

\newcommand{\inadom}[1]{\textsc{live}_{#1}}

\newcommand{\typename}[1]{\mathbf{#1}}
\newcommand{\types}{\mathfrak{D}}
\newcommand{\type}{\mathcal{D}}

\newcommand{\dom}{\Delta}
\newcommand{\sigp}{\Gamma}
\newcommand{\reals}{\mathbb{R}}
\newcommand{\naturals}{\mathbb{N}}
\newcommand{\ints}{\mathbb{Z}}
\newcommand{\strings}{\mathbb{S}}
\newcommand{\vartype}{\funsym{type}}

\newcommand{\ans}{\ensuremath{\mathit{ans}}\xspace}

\newcommand{\attrname}[1]{\ensuremath{\mathsf{#1}}}

\newcommand{\fod}[1][\types]{\textnormal{\texttt{FO}(\ensuremath{#1})}\xspace}
\newcommand{\qent}[3]{#1,#2\models #3}

\newcommand{\cset}{\mathcal{\E}}	
\newcommand{\pers}{\mathcal{P}}
\newcommand{\datq}[3]{#1(#2)\textsc{:-\,}#3}
\newcommand{\projindex}[2]{\relname{#1}[#2]}

\newcommand{\pk}[1]{\textsc{pk}(\relname{#1})}

\newcommand{\FKindex}[4]{\projindex{#1}{#2} \subseteq \projindex{#3}{#4}}

\newcommand{\PKindex}[2]{\pk{#1}=#2}

\newcommand{\cval}[1]{\mathsf{#1}}

\newcommand{\actions}{\mathcal{A}}
\newcommand{\queries}{\mathcal{Q}}
\newcommand{\dl}{\mathcal{L}}
\newcommand{\aop}[1]{\mathtt{#1}}
\newcommand{\aname}[1]{{#1}{\cdot}\aop{name}}
\newcommand{\apar}[1]{{#1}{\cdot}\aop{params}}
\newcommand{\aadd}[1]{{#1}{\cdot}\aop{add}}
\newcommand{\adel}[1]{{#1}{\cdot}\aop{del}}
\newcommand{\actname}[1]{\textsc{#1}}

\newcommand{\cpn}{\mathcal{N}\xspace}

\newcommand{\phigh}{{\color{blue}\cname{P\_High}}\xspace}
\newcommand{\plow}{{\color{blue}\cname{P\_Low}}\xspace}

\newcommand{\varsin}[1]{\setsym{Vars}(#1)}
\newcommand{\inflow}{F_{in}}
\newcommand{\outflow}{F_{out}}
\newcommand{\rbflow}{F_{rb}}

\newcommand{\colors}{\Sigma}

\newcommand{\transition}[1]{{\fontfamily{phv}\selectfont {#1}}}
\newcommand{\placename}[1]{\textit{#1}\xspace}

\newcommand{\places}{P}
\newcommand{\cplaces}{P_c}
\newcommand{\vplaces}{P_v}
\newcommand{\coloring}{\funsym{color}}
\newcommand{\quass}{\funsym{query}}
\newcommand{\guass}{\funsym{guard}}
\newcommand{\nuvarset}{\Upsilon_{\types}}
\newcommand{\vars}{\mathcal{X}_{\types}}

\newcommand{\transitions}{T}

\newcommand{\guards}[1]{\mathbb{F}_{#1}}

\newcommand{\tuples}[1]{\Omega_{#1}}

\newcommand{\invars}[1]{\setsym{InVars}(#1)}

\newcommand{\aass}{\funsym{act}}

\newcommand{\cl}{\mathcal{N}}

\newcommand{\net}{\mathcal{N}}
\newcommand{\varset}{\mathcal{V}_{\types}}

\newcommand{\tsys}[2]{\Gamma^{#1}_{#2}}
\newcommand{\states}{S}
\newcommand{\istate}{s_0}

\newcommand{\trans}{\rightarrow}

 \definecolor{dbcolor}{HTML}{F39C12}
\definecolor{datacolor}{HTML}{FAD7A0}
\definecolor{activedatacolor}{HTML}{85C1E9}
\definecolor{viewcolor}{HTML}{5499C7}
\definecolor{actioncolor}{HTML}{2471A3}
\definecolor{netcolor}{HTML}{D576AE}
\definecolor{mgreen}{rgb}{0.128,0.428,0}
\definecolor{burntorange}{rgb}{0.8, 0.33, 0.0}
\definecolor{camouflagegreen}{rgb}{0.47, 0.53, 0.42}
\definecolor{copperrose}{rgb}{0.6, 0.4, 0.4}

\tikzstyle{place}=[circle,thick,draw=black,fill=white,minimum size=7mm,font=\fontsize{9}{144}\selectfont]
\tikzstyle{transition}=[rectangle,thick,draw=black,fill=gray!20,minimum size=7mm]
\tikzstyle{enabledtransition}=[rectangle,very thick,draw=green!75,fill=green!20,minimum size=7mm]
 \tikzstyle{container}=[rectangle,rounded corners,very thick,draw=black!75,fill=black!20,minimum height=7mm,minimum width=14mm]
\tikzstyle{rplace}=[circle,ultra thick,draw=violet!75,fill=violet!20,minimum size=7mm]

\tikzstyle{erbox}=[draw, fill=gray!20, minimum width=7em, text width=6.0em, text centered,
  minimum height=3em,rounded corners,drop shadow]

\def\dbicon#1#2#3{
    \node at #1 [cylinder, shape border rotate=90, draw=white,fill=black,minimum height=#2,minimum width=#3,yshift=-.1cm] {};
}

\def\viewplace#1#2#3{
\begin{scope}[shift={#2}]
    \node (#1) at (0,0) [place,draw,label={#3}] {};
    \node at (0,0) [cylinder, shape border rotate=90, draw=white,fill=black,minimum height=.3cm,minimum width=.4cm,yshift=-.1cm] {};
\end{scope}
}

\tikzstyle{placelem}=[draw,
    cloud,
    cloud puffs = 15,
    minimum width=.2cm,
    minimum height=.2cm,
    fill=white,
    thick]

\tikzstyle{netelem}=[draw,
    cloud,
    cloud puffs = 20,
    minimum width=1.8cm,
    minimum height=1.8cm,
    fill=blue!10,
    ultra thick]

\tikzstyle{relationselem}=[placelem,fill=pink!20]
\tikzstyle{noopelem}=[placelem,fill=orange!20]
\tikzstyle{enteredplace}=[place,fill=yellow!20]
\tikzstyle{boundplace}=[place,fill=yellow!30]
\tikzstyle{guardokplace}=[place,fill=yellow!40]
\tikzstyle{updatedplace}=[place,fill=yellow!50]
\tikzstyle{violplace}=[place,fill=red!10]
\tikzstyle{constrokplace}=[place,fill=green!10]
\tikzstyle{docommitplace}=[place,fill=green!30]
\tikzstyle{dorollbackplace}=[place,fill=red!30]
\tikzstyle{arc}=[-stealth',thick]
\tikzstyle{readarc}=[-,thick]

\makeatletter
\pgfarrowsdeclare{X}{X}
{
  \pgfutil@tempdima=0.3pt%
  \advance\pgfutil@tempdima by.25\pgflinewidth%
  \pgfutil@tempdimb=5.5\pgfutil@tempdima\advance\pgfutil@tempdimb by.5\pgflinewidth%
  \pgfarrowsleftextend{+-\pgfutil@tempdimb}
  \pgfarrowsrightextend{0pt}
}
{
  \pgfutil@tempdima=0.3pt%
  \advance\pgfutil@tempdima by.25\pgflinewidth%
  \pgfsetdash{}{+0pt}
  \pgfsetroundcap
  \pgfsetmiterjoin
  \pgfpathmoveto{\pgfqpoint{-5.5\pgfutil@tempdima}{-6\pgfutil@tempdima}}
  \pgfpathlineto{\pgfqpoint{5.5\pgfutil@tempdima}{6\pgfutil@tempdima}}
  \pgfpathmoveto{\pgfqpoint{-5.5\pgfutil@tempdima}{6\pgfutil@tempdima}}
  \pgfpathlineto{\pgfqpoint{5.5\pgfutil@tempdima}{-6\pgfutil@tempdima}}
  \pgfusepathqstroke 
}
\makeatother

\makeatletter
\pgfarrowsdeclare{x}{x}
{
  \pgfutil@tempdima=0.3pt%
  \advance\pgfutil@tempdima by.25\pgflinewidth%
  \pgfutil@tempdimb=5.5\pgfutil@tempdima\advance\pgfutil@tempdimb by.5\pgflinewidth%
  \pgfarrowsleftextend{+-\pgfutil@tempdimb}
  \pgfarrowsrightextend{0pt}
}
{
  \pgfutil@tempdima=0.3pt%
  \advance\pgfutil@tempdima by.25\pgflinewidth%
  \pgfsetdash{}{+0pt}
  \pgfsetroundcap
  \pgfsetmiterjoin
  \pgfpathmoveto{\pgfqpoint{-3.5\pgfutil@tempdima}{-4\pgfutil@tempdima}}
  \pgfpathlineto{\pgfqpoint{3.5\pgfutil@tempdima}{4\pgfutil@tempdima}}
  \pgfpathmoveto{\pgfqpoint{-3.5\pgfutil@tempdima}{4\pgfutil@tempdima}}
  \pgfpathlineto{\pgfqpoint{3.5\pgfutil@tempdima}{-4\pgfutil@tempdima}}
  \pgfusepathqstroke 
}
\makeatother

\tikzset{
state/.style={
       rectangle split,
       rectangle split parts=2,
       rectangle split part fill={red!10,blue!10},
       rounded corners,
       draw=black,  thick,
       minimum height=2em,
       minimum width=2cm,
       inner sep=2pt,
       text centered,
       }
}

\usepackage{anyfontsize}
\usepackage{t1enc}
\colorlet{light-gray}{gray!20}
\definecolor{bronze}{rgb}{0.8, 0.5, 0.2}

\lstnewenvironment{SQL}
{\lstset{
   frameround=ffff,
    language=SQL,    
    numbers=none,
    breaklines=true,
    breakatwhitespace=true,
    mathescape=true,
    tabsize=3,
    keywordstyle=\color{black}\bfseries, 
    morekeywords={VALUES,SET,ENABLES},
    basicstyle=\small\ttfamily\color{black},
}}{}

\title{From DB-nets to\\ Coloured Petri Nets with Priorities\\ (Extended Version)}
\author{Marco Montali \and Andrey Rivkin}
\authorrunning{Montali, Rivkin}
\institute{
 Free University of Bozen-Bolzano,
 Piazza Domenicani 3, 39100 Bolzano, Italy\\
 \email{montali,rivkin@inf.unibz.it}
}

\begin{document}

\maketitle

\begin{abstract}
The recently introduced formalism of DB-nets has brought in a new conceptual way of modelling complex dynamic systems that equally account for the process and data dimensions, considering local data as well as persistent, transactional data. DB-nets combine a coloured variant of Petri nets with name creation and management (which we call $\nu$-CPN), with a relational database. The integration of these two components is realized by equipping the net with special ``view'' places that query the database and expose the resulting answers to the net, with actions that allow transitions to update the content of the database, and with special arcs capturing compensation in case of transaction failure. In this work, we study whether this sophisticated model can be encoded back into $\nu$-CPNs. In particular, we show that the meaningful fragment of DB-nets where database queries are expressed using unions of conjunctive queries with inequalities can be faithfully encoded into $\nu$-CPNs with transition priorities. This allows us to directly exploit state-of-the-art technologies such as CPN Tools to simulate and analyse this relevant class of DB-nets. (\textit{Topics covered: Higher-level net models, Relationships between Petri nets and other approaches})
\end{abstract}


\section{Introduction}

\definecolor{dbcolor}{HTML}{F39C12}
\definecolor{datacolor}{HTML}{FAD7A0}
\definecolor{activedatacolor}{HTML}{85C1E9}
\definecolor{viewcolor}{HTML}{5499C7}
\definecolor{actioncolor}{HTML}{2471A3}
\definecolor{netcolor}{HTML}{D576AE}

\begin{figure}[t]
\centering
\resizebox{.95\hsize}{!}{
\begin{tikzpicture}[x=1cm,y=1.5cm,>=triangle 60,thick]
\tikzstyle{elem} = [rectangle,rounded corners=5pt,minimum height=.7cm,minimum width=1.5cm,draw]
\tikzstyle{back} = [rectangle,rounded corners=10pt,minimum height=1.03cm,minimum width=9.3cm,anchor=west]

\node (db) at (0,-.25) [cylinder, shape border rotate=90, draw,minimum height=.9cm,minimum width=.9cm,left color=dbcolor!30,right color=dbcolor]
{~~~~};
\node (dblegend) at (.6,-.2) [anchor=west] {DB};

\node (action) at(1,1) [elem,left color=actioncolor!30, right color=actioncolor] {Actions};
\node (view) at(-1.3,1) [elem,left color=viewcolor!30, right color=viewcolor] {Queries};


\viewplace{viewplace}{(-1.3,2.1)}{[yshift=.01cm,xshift=-.5cm]View places}
\draw[X-stealth'] (-.4,1.9) to node[anchor=west,xshift=.2cm]{} (0.4,1.9);
\node (rlalegend) at (-.1,2.38) {$\begin{array}{@{}c@{}} \text{Rollback} \\ \text{arcs}\end{array}$};
\node (atransition) at (1,2.1) [transition] {$\action$};
\node (tlegend) at (1.05,2.55) [anchor=west,xshift=-.4cm] {Transitions};
\node (transition) at (2.2,2.1) [transition] {};
\node (place) at (3.35,2.1) [place] {};
\node (plegend) at (3.35,2.55) {Places};
\node (nuvar) at (5.12,2.4) {${\Large\nu}$\; Fresh vars.};
\draw[-stealth'] (4,2.15) to node[anchor=west,xshift=.2cm]{Arcs}  (4.5,2.15);
\draw[-] (4,1.9) to node[anchor=west,xshift=.2cm]{Read arcs} (4.5,1.9);

\draw[->] (db) to node[anchor=east]{fetch} (view);
\draw[->] (action) to node[anchor=west]{update} (db);
\draw[->] (view) to node[anchor=east,yshift=-1mm]{populate} (viewplace);
\draw[->] (atransition) to node[anchor=west,yshift=-1mm]{trigger} (action);

\node[draw=black, fill=white] (dbext) at (-2.4,2.95) {relational part};
\node[draw=black, fill=white] (dbext) at (5.7,2.95) {\nucpn part};

  \begin{pgfonlayer}{background}
\node (data) at (-3,-.05) [back,fill=datacolor] {};
\node (activedata) at (-3,1) [back,fill=activedatacolor] {};
\node (netdata) at (-3,2.18) [back,fill=netcolor,minimum height=1.7cm] {};
\node (datalegend) at (-3,-.2) [anchor=east] {\bf persistence layer};
\node (datalegend2) at (3,-.2) [anchor=west] {\phantom{\bf persistence layer}};
\node (activedatalegend) at (-3,1) [anchor=east] {\bf data logic layer};
\node (netlegend) at (-3,2.18) [anchor=east] {\bf control layer};

\filldraw[draw=black,dashed,fill=none,line width=0.2mm] (-2.8,1.58) rectangle (1.574,2.9);
\filldraw[draw=black,dashed,fill=none,line width=0.2mm] (1.78,1.58) rectangle (6.16,2.9);

 \end{pgfonlayer}

\end{tikzpicture}
}
\caption{The conceptual components of \dbnets\label{fig:dbnets}}
\end{figure}
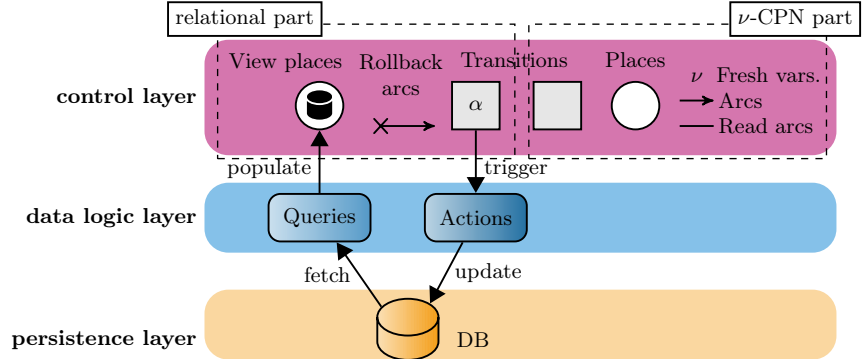

During the last decade, the Business Process Management (BPM) community has gradually lifted its attention from process models mainly focusing on the flow of activities to multi-perspective models that also account for the interplay between the process and the data perspective \cite{Hull08,Reic12,CaDM13}. In particular, several variants of high-level Petri nets have been adopted to capture meaningful integrated models for processes and data, at the same time retaining the possibility of analysing the resulting state space (see, e.g., \cite{MonR16,TriS16,deLFM18,MonR17}). 

In this spectrum, the recently introduced formalism of DB-nets \cite {MonR17} has brought in a new conceptual way of modelling complex dynamic systems that equally account for the process and data dimensions, considering local data as well as persistent, transactional data. On the one hand, a DB-net adopts a standard relational database with constraints to store persistent data. The database can be queried through SQL/first-order queries, and updated via actions in a transactional way (that is, committing the update only if the resulting database satisfies all intended constraints). On the other hand, a DB-net employs a coloured variant of a Petri net with name creation and management \cite{RVFE11} to capture the process control-flow, the injection of (possibly fresh) data such as the creation of new case identifiers \cite{MonR16}, and tuples of typed data locally carried out by tokens. This model, which we call $\nu$-CPN, can be seen as a fragment of standard Coloured Petri nets \cite{JeK09} with pattern matching on inscriptions, infinite colour domains, boolean guards, and a very limited use of SML to account for fresh data injection. This also means that $\nu$-CPNs can be seamlessly modelled, simulated, and analysed using state-of-the-art tools such as CPN Tools. 

The integration of these two components is realized in a DB-net by extending the $\nu$-CPN with three novel constructs:
\begin{inparaenum}[\it (i)]
\item \emph{view places}, special places that query the database and expose the resulting answers as coloured tokens that can be inspected but not directly consumed;
\item \emph{action bindings}, linking transitions to database updates by mapping inscription variables to action parameters;
\item \emph{rollback transition-place arcs}, capturing the emission of tokens in case a fired transition induces a  failing database update, and in turn supporting the enablement of compensation transitions.
\end{inparaenum}
All conceptual components used in the DB-net model are depicted in Figure~\ref{fig:dbnets}. Notably, DB-nets have been employed to formalize application integration patterns \cite{RMRS18}.

In this work, we study whether this sophisticated model can be \emph{encoded back into $\nu$-CPNs}, with a twofold intention. On the foundational side, we aim at understanding whether the process-data integration realized in DB-nets adds expressiveness to $\nu$-CPNs, or it is instead conceptual, syntactic sugar. On the practical side, the existence of an encoding would allow us to directly exploit state-of-the-art tools such as CPN Tools towards simulation and analysis of DB-nets. In the case of CPN Tools, this is the only way possible when it comes to state space construction, given the fact that this feature cannot be refined through the third-party extension mechanism offered by the framework.

Specifically, we constructively show through a behavior-preserving translation mechanism that this encoding is indeed possible for a large and meaningful class of DB-nets, provided that the obtained $\nu$-CPN is equipped with \emph{transition priorities}~\cite{WestergaardV11} (a feature that is supported by virtually all CPN frameworks, including CPN Tools). Such class corresponds to DB-nets where the database is equipped with key, foreign key, and domain constraints, and where the view places query the database using unions of conjunctive queries (UCQs) with inequalities. Such query language corresponds to the widely adopted fragment of SQL consisting of select-project-join queries with filters \cite{AbHV95}. 

\section{The \dbnet Formal Model}

In this section, we briefly present the key concepts and notions used for defining \dbnets. Conceptually, a \dbnet is composed of three layers (cf. Figure~\ref{fig:dbnets})
\begin{inparaenum}[1)]
\item \emph{persistence layer}, capturing a full-fledged relational database with constraints,
and used to store background data, and data that are persistent across cases;
\item \emph{control layer}, employing a variant of CPNs to capture the process control-flow,
case data, and possibly the resources involved in the process execution;
\item \emph{data logic layer}, interconnecting in the persistence and the control layer.
\end{inparaenum}
\begin{definition}
A \emph{db-net} is a tuple $\tup{\types,\pers,\dl,\cl}$, where:
\begin{inparaenum}[\it (i)]
\item $\types$ is a type domain;
\item $\pers$ is a $\types$-typed persistence layer;
\item $\dl$ is a $\types$-typed data logic layer over $\pers$;
\item $\cl$ is a $\types$-typed control layer over $\dl$.
\end{inparaenum}
\end{definition}
We next formalize the framework layer by layer.

\smallskip\noindent
\textbf{Persistence layer.} A \emph{type domain} $\types$ is a finite set of pairwise disjoint \emph{data types}$\type=\tup{\dom_\type,\sigp_\type}$, where $\dom_\type$ is a \emph{value domain}, and $\sigp_\type$ is a finite set of \emph{predicate  symbols}. Examples of data types are:
\begin{inparaenum}[\it (i)]
\item $\typename{string} : \tup{\strings,\set{=_s}}$, strings with the equality predicate;
\item $\typename{real} : \tup{\reals,\set{=_r,<_r}}$, reals with the usual comparison operators; 
\item $\typename{int} : \tup{\ints,\set{=_{int},<_{int},succ}}$, integers with the usual comparison operators, as well as the successor predicate.
\end{inparaenum}

A \emph{$\types$-typed database schema} $\schema$ is a finite set of $\types$-typed relation schemas $R(\type_1,\ldots,\type_n)$, where $\type_i$ indicates the data type associated to an $i$-th component of $R$. 
A \emph{$\types$-typed database instance $\I$ over $\schema$} is a finite set of facts of the form $R(\cname{o_1},\ldots,\cname{o_n})$, such that 
$R(\type_1,\ldots,\type_n) \in \schema$ and 
$\cname{o_i}\in \dom_{\type_i}$, for $i \in \set{1,\ldots,n}$.
Given a type $\type \in \types$, the \emph{$\type$-active domain of $\I$}, is the set of $\adom[\type]{\I}=\set{\cname{o} \in \dom_\type\mid \cname{o} \text{ occurs in }\I}$. 

Given a type domain $\types$, we fix a countably infinite set $\varset$ of typed variables with a \emph{variable typing function} $\vartype: \varset \rightarrow \types$. 
As a query language, we adopt standard first-order logic (FOL) extended with data types under the active-domain semantics~\cite{Lib04}, that is, the evaluation of quantifiers only depends on the values explicitly appearing in the database instance over which they are applied. This can be seen as the FOL representation of SQL queries. A\emph{(well-typed) \fod query} $Q$ over a $\types$-typed database schema $\schema$ has the form $\set{\vec{x}\mid \varphi(\vec{x})}$, where $\vec{x}$ is the tuple of answer variables of $Q$, and $\varphi$ is a FO formula, with $\vec{x}$ as free variables, over predicates in $\cup_{\type\in\types}\sigp_\type$ and relation schemas in $\schema$, whose variables and constants are correctly typed.
We use $Q(\vec{x})$ to make the answer variables $\vec{x}$ of $Q$ explicit, and denote the set of such variables as $\free{Q}$. When $\free{Q}=\emptyset$, we call $Q$ a \emph{boolean query}.

A \emph{substitution} for a set $X = \set{x_1,\ldots,x_n}$ of typed variables,
is a function $\subst: X \rightarrow \dom_\types$, such that
$\subst(x) \in \dom_{\vartype(x)}$, for every $x \in X$.  A \emph{substitution
 $\subst$ for a \fod query $Q$} is a substitution for the free variables of
$Q$. We denote by $Q\subst$ the boolean query obtained from $Q$ by replacing
each occurrence of a free variable $x \in \free{Q}$ with the value $\subst(x)$.
Given a $\types$-typed database schema $\schema$, a $\types$-typed instance
$\I$ over $\schema$, and a \fod query $Q$ over $\schema$, the set of
\emph{answers} to $Q$ in $\I$ is defined as the set
$\ans(Q,\I) = \{ \subst: \free{Q} \rightarrow \adom[\types]{\I} \mid
\qent{\I}{\subst}{Q}\}$ of substitutions for $Q$, where $\models$ denotes
standard FO entailment (i.e., we use \emph{active-domain semantics}).
We denote by $\inadom{\type}(x)$ the unary query returning all the objects of
type $\type$ that occur in the active domain (writing such a query is
straightforward).
When $Q$ is boolean, we write $\ans(Q,\I) \equiv \true$ if $\ans(Q,\I)$
consists only of the empty substitution (denoted $\tup{}$),
and $\ans(Q,\I) \equiv \false$ if $\ans(Q,\I) = \emptyset$.
Boolean queries are also used to express \emph{constraints} over $\schema$.  
We introduce
explicitly two common types of constraints: given relations $R/n$ and $S/m$,
and two index-sets $N$ and $M$ such that $1\leq i\leq n$ for every $i\in N$,
and $1\leq j\leq m$ for every $j \in M$, we fix the following notation:
\begin{inparaenum}[\it (i)]
\item  $\PKindex{R}{N}$ expresses that the projection
  $\projindex{R}{N}$ of $R$ on $N$ is a primary key for $R$;
\item $\FKindex{R}{N}{S}{M}$ expresses that the projection $\projindex{R}{N}$ of $R$ on
  $N$ refers the projection $\projindex{S}{M}$ of $S$ on $M$, which has to be a key
  for $S$.
\end{inparaenum}
Both kinds of constraints are obviously expressible as suitable queries~\cite{AbHV95}.

%
\begin{definition}
A $\types$-typed \emph{persistence layer} is a pair $\tup{\schema,\cset}$ where:
\begin{inparaenum}[\it (i)]
\item $\schema$ is a $\types$-typed database schema;
\item $\cset$ is a finite set $\{\Phi_1,...,\Phi_k\}$ of boolean \fod queries over $\schema$, modelling \emph{constraints over $\schema$}.
\end{inparaenum}
\end{definition}
We say that a $\types$-typed database instance $\I$ \emph{complies with} $\pers$, if $\I$ is defined over $\schema$ and satisfies all constraints in $\cset$.

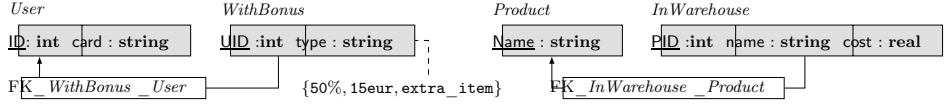
\begin{figure}[t!]
\centering
\resizebox{1.02\textwidth}{!}{
\begin{tikzpicture}[yshift =-5.4cm,xshift=-1cm,relation/.style={rectangle split, rectangle split parts=#1, rectangle split part align=base, draw, anchor=center, align=center, text height=3mm, text centered}]\hspace*{-0.3cm}


\node (User_title) {\relname{User}};
\node [relation=2, rectangle split horizontal, rectangle split part fill={lightgray!50}, below=0.6cm of User_title.west, anchor=west] (User)
{\underline{\attrname{ID}}: $\typename{int}$ 
\nodepart{two} \attrname{card} : $\typename{string}$};

\node [right=4cm of User_title.west, anchor=west] (WithB_title) {\relname{WithBonus}};
\node [relation=2, rectangle split horizontal, rectangle split part fill={lightgray!50}, anchor=north west, below=0.6cm of WithB_title.west, anchor=west] (WithB)
{\underline{\attrname{UID}} :$\typename{int}$%
\nodepart{two} \attrname{type} : $\typename{string}$};

\node [right=3.3cm of WithB_title.east, anchor=west] (Product_title) {\relname{Product}};
\node [relation=1, rectangle split horizontal, rectangle split part fill={lightgray!50}, anchor=north west, below=0.6cm of Product_title.west, anchor=west] (Product)
{\underline{\attrname{Name}} : $\typename{string}$};

\node [right=3cm of Product_title.west, anchor=west] (Warehouse_title) {\relname{InWarehouse}};
\node [relation=3, rectangle split horizontal, rectangle split part fill={lightgray!50}, anchor=north west, below=0.6cm of Warehouse_title.west, anchor=west] (Warehouse)
{\underline{\attrname{PID}} :$\typename{int}$%
\nodepart{two} \attrname{name} : $\typename{string}$
\nodepart{three}  \attrname{cost} : $\typename{real}$};

\node[below=0.35cm of WithB.south,xshift=1.8cm] (values) {$\set{\cname{50\%},\cname{15eur},\cname{extra\_item}}$};


\draw[-latex] ($(WithB.one south) +(0.2,0)$) -- ++ (0,-0.60) -| node[rectangle,draw=black,fill=white,xshift=14mm,inner sep=.5mm,minimum size=1mm]{\footnotesize{\textup{FK\_\relname{WithBonus}\_\relname{User}}}} ($(User.one south) + (-0.2,0)$);
\draw[-latex] ($(Warehouse.two south) +(0.2,0)$) -- ++ (0,-0.60) -| node[rectangle,draw=black,fill=white,xshift=23mm,inner sep=.5mm,minimum size=1mm]{\footnotesize{\textup{FK\_\relname{InWarehouse}\_\relname{Product}}}} ($(Product.one south) + (-0.2,0)$);
\draw[dashed] ($(values.north)+(0.2,0)$) --++ (0,.65) -| (WithB.two east) ;

\end{tikzpicture}
}
\caption{The persistence layer for the online shopping scenario} \label{fig:shopping-cart-db}
\end{figure}

\begin{example}\label{ex:pl}
Let us consider a simplified shopping process used by an e-commerce website. Specifically, we are interested in a simplified scenario in which an already registered user logs in the website and immediately proceeds with selecting  products. While products can be selected and added to the shopping cart, the user can occasionally choose a monthly bonus that may be applied when concluding a purchase. We restrict this scenario only by considering cases in which each user ends up buying at least one product. 

The persistence layer $\pers=\tup{\schema,\cset}$ of this scenario comprises four relation schemas (cf. Figure~\ref{fig:shopping-cart-db}): $\relname{User}(\typename{int},\typename{string})$ lists registered users together with their credit card data, $\relname{WithBonus}(\typename{int},\typename{string})$ indicates users that have bonuses,  $\relname{Product}(\typename{string})$ indexes product types offered by the website and $\relname{InWarehouse}(\typename{int}, \typename{string},\typename{real})$ models products (together with their costs) stored in the warehouse. Note the constraints between these schemas. 
For example, in order to show that users cannot have more than one bonus at a time, we introduce a foreign key constraint between $\relname{WithBonus}$ and $\relname{User}$ that is denoted as $\FKindex{WithBonus}{\set{1}}{User}{\set{1}}$ and formalized in FO logic as: $\forall uid,bt. \relname{WithBonus}(uid,bt)\rightarrow\exists card. \relname{User}(uid,card)$. 
Another constraint limits the bonus type values in $\relname{WithBonus}$ and can be expressed as $\forall uid,bt. \relname{WithBonus}(uid,bt)\rightarrow bt=\cname{50\%}\lor bt=\cname{15eur}\lor bt=\cname{extra\_item}$.\qed
\end{example}


\noindent
\textbf{Data logic layer.}
The data logic layer allows one to \emph{extract} data from the database instance using queries as well as to \emph{update} the database instance by adding and deleting possibly multiple facts at once. The updates follow the \emph{transactional} semantics: if a new database instance obtained after some update is still compliant with the persistence layer, the update is \emph{committed}; otherwise it is \emph{rolled back}. Such updates are realized in parametric atomic actions, resembling ADL actions in planning~\cite{DrTh08}, and consist of fact templates -- expressions that, once instantiated, assert which facts will be added to and deleted from the database. 
Specifically, given a typed relation $R(\type_1,\ldots,\type_n) \in \schema$, an $R$-fact template over $\vec{p}$ 
has the form $R(y_1,\ldots,y_n)$, such that for every $i \in \set{1,\ldots,n}$, $y_i$ is either a value $\cname{o} \in \dom_{\type_i}$, or a variable $x \in \vec{p}$ with $\vartype(x) = \type_i$.
 

A \emph{(parameterized) action} over a $\types$-typed persistence layer $\tup{\schema,\cset}$ is a tuple $\tup{\cname{n},\vec{p},F^+,F^-}$, where:
\begin{inparaenum}[\it (i)]
	\item $\cname{n}$ is the \emph{action name};
	\item $\vec{p}$ is a tuple of pairwise distinct variables from $\varset$, denoting the \emph{action (formal) parameters};
	\item $F^+$ and $F^-$ respectively represent a finite set of $\schema$-fact templates (i.e., some $R$-fact templates for some $R\in\schema$)  over $\vec{p}$, to be \emph{added} to and \emph{deleted} from the current database instance.
\end{inparaenum}
To access the different components of an action $\action$, we use a dot notation: $\aname{\action} = \cname{n}$, $\apar{\action} = \vec{p}$, $\aadd{\action}= F^+$, and $\adel{\action}=F^-$.
Given an action $\action$ and a (parameter) substitution $\subst$ for $\apar{\action}$, we call \emph{action instance $\action\subst$} the (ground) action resulting by substituting parameters of $\action$ with corresponding values from $\subst$. Then, given a $\types$-typed database instance $\I$ compliant with $\types$, the \emph{application} of $\action\subst$ on $\I$, , written $\doact{\action\subst}{\I}$, is a database instance over $\schema$ obtained as $(\I\setminus F^-_{\action\subst})\cup F^+_{\action\subst}$, where:
\begin{inparaenum}[\it (i)]
\item $F^-_{\action\subst} = \bigcup_{R(\vec{y}) \in \adel{\action}} R(\vec{y})\subst$;
\item $F^+_{\action\subst} = \bigcup_{R(\vec{y}) \in \aadd{\action}} R(\vec{y})\subst$.
\end{inparaenum}
If $\doact{\action\subst}{\I}$ complies with $\pers$, $\action\subst$ can be \emph{successfully applied} to $\I$. 
Note that, in order to avoid situations in which the same fact is asserted to be added and deleted, we prioritize deletions over additions.

\begin{definition}
	Given a $\types$-typed persistence layer $\pers$, a \emph{$\types$-typed data logic layer over $\pers$} is a pair $\tup{\queries,\actions}$, where:
	\begin{inparaenum}[\it (i)]
	\item $\queries$ is a finite set of \fod queries over $\pers$;
	\item $\actions$ is a finite set of actions 	over $\pers$.
	\end{inparaenum}
\end{definition}

\begin{example}\label{ex:dl}
We make the scenario of Example~\ref{ex:pl} operational, introducing a data logic layer $\dl$ over $\pers$. To inspect the persistence layer, we use the following queries:
\begin{compactitem}[$\bullet$]
\item $\datq{\cname{Q_{products}}}{pid,n,c}{\relname{Product}(n)\land\relname{InWarehouse}(pid,n,c)\land c\neq\cname{null}}$, to extract products available in the warehouse and whose price is not $\cname{null}$ (those without prices can be undergoing the stock-taking process);
\item $\datq{\cname{Q_{users}}}{uid}{\exists card.\relname{User}(id,card)}$, to get registered users;
\item $\datq{\cname{Q_{wbonus}}}{uid,bt',u}{\relname{WithBonus}(uid,bt',u)}$, to inspect all users with bonuses.
\end{compactitem}

In addition, $\dl$ provides key functionalities for organizing the shopping process.
Such functionalities are realized through four actions (where, for simplicity, we blur the distinction between an action and its name).
To manage bonuses we use two actions $\actname{addb}$ and $\actname{change}$. The former is used to assign a bonus of type $bt$ to a user with id $uid$ ($\apar{\actname{addb}} = \tup{uid,bt}$) and record it into the persistent storage:
$\aadd{\actname{addb}}= \set{\relname{WithBonus}(uid,bt)}$, $\adel{\actname{addb}}=\emptyset$.
Note that, before logging in, the user may have already a bonus assigned during one of the previous sessions. At will, such a bonus can be changed using the following action: $\aadd{\actname{change}}= \set{\relname{WithBonus}(uid,bt')}$, $\adel{\actname{change}}=\set{\relname{WithBonus}(uid,bt)}$.
In fact, $\actname{change}$ realizes an update by first deleting a tuple that is characterized by $uid$ and $bt'$ (the old bonus),  and then adding its modified version. 
We use $\actname{reserve}$ to reserve product $pid$ stored in cart $cid$ for further processing (e.g., the preparation for shipment) by deleting it from the list of available products: $\aadd{\actname{reserve}}= \emptyset$, $\adel{\actname{reserve}}=\set{\relname{InWarehouse}(pid,n,c)}$.
At last, we may utilize our monthly bonus (if it has not been yet used) to consider it when paying the order. For that, we use an action called $\actname{apply}$ such that: $\aadd{\actname{apply}}= \emptyset$, $\adel{\actname{apply}}=\set{\relname{WithBonus}(uid,bt)}$. \qed
\end{example}

\noindent
\textbf{Control layer.} The control layer employs a fragment of Coloured Petri net to capture the process control flow and a data logic to interact with an underlying persistence layer.
We fix some preliminary notions. We consider the standard notion of a \emph{multiset}. Given a set $A$,  the \emph{set of multisets} over $A$, written $\mult{A}$, is the set of mappings of the form $m:A\rightarrow \mathbb{N}$.
Given a multiset $S \in \mult{A}$ and an element $a \in A$, $S(a) \in \mathbb{N}$ denotes the number of times $a$ appears in $S$.  Given $a \in A$ and $n \in \mathbb{N}$, we write $a^n \in S$ if $S(a) = n$.  We also consider the usual operations on multisets. Given $S_1,S_2 \in \mult{A}$:
\begin{inparaenum}[\it (i)]
\item $S_1 \subseteq S_2$ (resp., $S_1 \subset S_2$) if $S_1(a) \leq S_2(a)$ (resp., $S_1(a) < S_2(a)$) for each $a \in A$;
\item $S_1 + S_2 = \set{a^n \mid a \in A \text{ and } n = S_1(a) + S_2(a)}$;
\item if $S_1 \subseteq S_2$, $S_2 - S_1 = \set{a^n \mid a \in A \text{ and } n = S_2(a) - S_1(a)}$;
\item given a number $k \in \mathbb{N}$, $k \cdot S_1 = \set{a^{kn} \mid a^n \in S_1}$.
\end{inparaenum}

We shall call \emph{inscription} a tuple of typed variables (and, possibly, values) and denote the set of all possible inscriptions over set $\mathcal{Y}$ as $\tuples{\mathcal{Y}}$, and the set of variables appearing inside an inscription $\omega \in \tuples{\mathcal{Y}}$ as $\varsin{\omega}$ (such notation naturally extends to sets and multisets of inscriptions).  
In the spirit of CPNs, the control layer assigns to each place a color type, which in turn combines one or more data types from $\types$. Formally, a \emph{$\types$-color} is $\type_1 \times \ldots \times \type_m$, where for each $i \in \set{1,\ldots,m}$, we have $\type_i \in \types$. We denote by $\colors$ the set of all possible $\types$-colors.
To account for fresh external inputs, we employ the well-known mechanism adopted in $\nu$-Petri nets~\cite{RVFE11,MonR16} and introduce a countably infinite set $\nuvarset$ of $\types$-typed \emph{fresh variables}. To guarantee an unlimited provisioning of fresh values, we impose that for every variable $\nu \in \nuvarset$, we have that $\dom_{\vartype(\nu)}$ is countably infinite. Hereinafter, we shall fix a countably infinite set of $\types$-typed variable $\vars = \varset \cup \nuvarset$ as the disjoint union of ``normal" variables $\varset$ and fresh variables $\nuvarset$.

As we have mentioned before, the control layer can be split into two parts. 
Let us first define the \nucpn part that can be seen as an extension of $\nu$-Petri nets with concrete data types, boolean (type-aware) guards and read arcs. 

\begin{definition}
A $\types$-typed \nucpn $\cpn$ is a tuple $\tup{\places,\transitions,\inflow,\outflow,\coloring}$, where:
\begin{compactenum}
\item $\places$ is a finite set of places.
\item $\coloring: \places \rightarrow \colors$ is a color type assignment over $\places$ mapping each place $p \in \places$ to a corresponding $\types$-type color.
\item $\transitions$ is a finite set of transitions, such that $\transitions \cap \places= \emptyset$.
\item $\inflow: \places \times \transitions \rightarrow \mult{\tuples{\varset}}$ is an input flow from $\places$ to $\transitions$ assigning multisets of inscriptions (over variables $\varset$) to input arcs, s.t. that each of such inscriptions $\tup{x_1,\ldots,x_m}$ is compatible with each of its input places $p$, i.e., for every $i \in \set{1,\ldots,m}$, we have $\vartype(x_i) = \type_i$, where $\coloring(p) = \type_1 \times \ldots\times \type_m$.
\item $\guass:  \transitions \rightarrow \guards{\types}$ is a transition guard assignment over $\transitions$ 
assigning to each transition $t\in\transitions$ a $\types$-typed guard $\varphi$, s.t.:
\begin{compactitem}[$\bullet$]
\item $\invars{t} = \set{x \in \varset \mid \text{there exists }p \in \places \text{ such that }x \in \varsin{\inflow(\tup{p,t})}}$ is the set of all variables occurring on input arc inscriptions of $t$;
\item a $\types$-typed \emph{guard} from is a formula (or a quantifier- and relation-free \fod query) of the form $\varphi~::=~\true \mid S(\vec{y}) \mid \neg \varphi  \mid \varphi_1 \land \varphi_2$, where $S/n\in\sigp_\type$ and, for $\vec{y} = \tup{y_1,\ldots,y_n} \subseteq \varset$, we have that  $y_i$ is either a value $\cname{o} \in \dom_{\type}$, or a variable $x_i \in \varset$ with $\vartype(x_i) = \type$ ($i \in \set{1,\ldots,n}$);
\item $\guards{\types}$ is the set of all possible $\types$-typed guards and, with a slight abuse of notation, $\varsin{\varphi}$ is the set of variables occurring in $\varphi$.
\end{compactitem}
\item $\outflow: \transitions \times \places \rightarrow \mult{\tuples{\vars \cup \dom_\types}}$ is an \emph{output flow} from transitions $\transitions$ to  places $\places$  assigning multisets of inscriptions to output arcs, such that all such inscriptions are compatible with their output places.
\end{compactenum}
\end{definition}

According to the diagram in Figure~\ref{fig:dbnets}, the \dbnet control layer can be obtained on top of \nucpns by essentially adding three mechanisms that allow the net to interact with the underlying persistent storage:
\begin{inparaenum}[\it (i)]
\item view places, allowing the net to inspect parts of the database using queries;
\item action binding, linking atomic actions and their parameters to transitions and their inscription variables;
\item rollback transition-place arcs, enacted when the action application induced by a transition firing violates some database constraint, so as to explicitly account for ``error-handling''.
\end{inparaenum}

\begin{definition}
A $\types$-typed \emph{control layer} over a data logic layer $\dl = \tup{\queries,\actions}$ is a tuple
$\tup{\places,\transitions,\inflow,\outflow,\rbflow,\coloring,\quass,\guass,\aass}$, where:
\begin{compactenum}

\item $\tup{\cplaces,\transitions,\inflow,\outflow,\coloring}$ is a  $\types$-typed \nucpn, where $\cplaces$ is a finite set of control places.
\item $\places = \cplaces \cup \vplaces$ is a finite set of places, where $\vplaces$ are view places (decorated as \resizebox{.35cm}{!}{\tikz{\node [circle,draw,very thick,minimum width=.6cm, minimum height=.6cm] at (0,0) {}; \dbicon{(0,0)}{.4cm}{.3cm}}} and connected to transitions with special read arcs).
\item $\quass: \vplaces \rightarrow \queries$ is a query assignment mapping each view place $p \in \vplaces$ with $\coloring(p) = \type_1\times\ldots\times\type_n$ to a query $Q(x_1,\ldots,x_n)$ from $\queries$, s.t. the color of $p$ component-wise matches with the types of the free variables in $Q$: for each $i \in \set{1,\ldots,n}$, we have $\type_i = \vartype(x_i)$.
\item $\aass: \transitions \rightarrow \actions \times \tuples{\vars \cup \dom_\types}$ is a partial function assigning transitions in $\transitions$ to actions in $\actions$, where $\aass(t)$ maps $t$ to an action $\action \in \actions$ together with a (binding) inscription $\tup{y_1,\ldots,y_m}$, s.t. if $\apar{\action} = \tup{z_1,\ldots,z_m}$ and, for each $i \in \set{1,\ldots,m}$, we have $\vartype(y_i) = \vartype(z_i)$ if $y_i$ is a variable from $\vars$, or $y_i \in \dom_{\vartype(z_i)}$ if $y_i$ is a value from $\dom_\types$. 
\item $\rbflow$ is an output flow from $\transitions$ to $\cplaces$ called \emph{rollback flow} (we shall refer to $\outflow$ as \emph{normal output flow}).
\end{compactenum}
\end{definition}

\begin{figure}[t!]
\centering
\resizebox{.90\hsize}{!}{
\begin{tikzpicture}[->,>=stealth',auto,x=2.2cm,y=1.7 cm,thick]
  \viewplace{users}{(0,-.7)}{below,xshift=.0cm:\emph{Users}};
  \viewplace{products}{(0,1.3)}{below,xshift=.0cm:\emph{Products}};
  \viewplace{bonuses}{(5.2,-2)}{below,xshift=.0cm:$\begin{array}{@{}c@{}} \text{\emph{Bonus}} \\ \text{\emph{Holders}}\end{array}$};


  \node[transition, label=left:$\begin{array}{@{}c@{}} \text{\transition{Log}} \\ \text{\transition{In}}\end{array}$] (login) at (0,0) {};
  
  \node[place] (logged) at (1,0) {};
  
  \node[transition, label=below:$\begin{array}{@{}c@{}} \text{\transition{Add}} \\ \text{\transition{Product}}\end{array}$] (addprod) at (1,1.3) {};
  
  \node[transition, label=below:$\begin{array}{@{}c@{}} \text{\transition{Acquire}} \\ \text{\transition{Bonus}}\end{array}$] (acqbonus) at (1,-2) 
  {$\begin{array}{@{}c@{}}
                 \text{\actname{addb}}\\
  				(uid,bt)
               \end{array}
               $ };
	\node[place,label=below:Cart] (cart) at (2.28,1.3) {};
	
	\node[place] (rebonus) at (2.35,-2) {};
	
	 \node[transition, label=below:$\begin{array}{@{}c@{}} \text{\transition{Change}} \\ \text{\transition{Bonus}}\end{array}$] (changebonus) at (3.9,-2) 
  {$\begin{array}{@{}c@{}}
                 \text{\actname{change}}\\
  				(uid,bt,bt',u)
               \end{array}
               $ };

  \node[transition, label=below:$\begin{array}{@{}c@{}} \text{\transition{Check Out}}\end{array}$] (checkout) at (2.6,0) {};
  
    \node[transition, label=left:$\begin{array}{@{}c@{}} \text{\transition{Keep}} \\ \text{\transition{Bonus}}\end{array}$] (keep) at (2.35,-1.2) {};
  
  	\node[place] (checked) at (3.7,0) {};
  	\node[transition, label=below:$\begin{array}{@{}c@{}} \text{\transition{Prepare}} \\ \text{\transition{Order}}\end{array}$] (prepare) at (3.7,1.3) 
  	{$\begin{array}{@{}c@{}}
                 \text{\actname{reserve}}\\
  				(pid, n, c)
               \end{array}
               $ };
    \node[place,label=below:Reserved] (reserved) at (5.1,1.3) {};
               
       	\node[transition, label=below:$\begin{array}{@{}c@{}} \text{\transition{Finish Order}}\end{array}$] (pay) at (4.9,0){};
       		\node[transition, label=below:$\begin{array}{@{}c@{}} \text{\transition{Finish Order}} \\ \text{\transition{With Bonus}}\end{array}$] (paybonus) at (6.3,-2) 
  	{$\begin{array}{@{}c@{}}
                 \text{\actname{apply}}\\
  				(uid,bt)
               \end{array}
               $ };

           \node[
    draw,
    cloud,
    cloud puffs = 15,
    minimum width=1cm,
    minimum height=0.75cm
	](done) at (6.3,0) {\Large{...}};

\path
(login) edge node[above] {$\tup{uid,\nu_{c}}$}  (logged)
(logged) edge[bend left=40] node[left,pos=0.5] {$\tup{uid,cid}$}  (addprod.south west)
(addprod.south east) edge[bend left=40] node[right,pos=0.5] {$\tup{uid,cid}$}  (logged)
(logged) edge[bend left=0] node[right,pos=0.8] {$\tup{uid,cid}$}  (acqbonus.north)
(acqbonus.north west) edge[bend left=30] node[left,pos=0.2] {$\tup{uid,cid}$}  (logged)
(rebonus) edge node[below,pos=0.5] {$\tup{uid,cid,bt}$} (changebonus)
(rebonus) edge node[right,pos=0.5] {$\tup{uid,cid}$} (keep)
(keep.north west) edge node[below,sloped,pos=0.6] {$\tup{uid,cid}$}  (logged.south east)
(changebonus.north) edge node[sloped,above,pos=0.4] {$\tup{uid,cid}$} (logged)
(addprod) edge node[below] {$\tup{cid,pid,n,c}$}  (cart)
(logged) edge node[above] {$\tup{uid,cid}$}  (checkout)
(checkout) edge node[above] {$\tup{uid,cid}$} (checked)
(cart) edge node[below] {$\tup{cid,pid,n,c}$} (prepare)
(checked) edge[bend left=30] node[left,pos=0.7] {$\tup{uid,cid}$} (prepare.south west)
(prepare.south east) edge[bend left=30] node[right,pos=0.3] {$\tup{uid,cid}$} (checked)
(prepare) edge node[below,pos=0.5] {$\tup{cid,pid,c}$} (reserved)
(checked) edge node[above] {$\tup{uid,cid}$} (pay)
(checked) edge[bend right=0] node[above,sloped,pos=0.6] {$\tup{uid,cid}$} (paybonus.north)
(paybonus) edge node[above,sloped] {$\tup{uid,cid,dest}$} (done)
(pay) edge node[above] {$\tup{uid,cid,dest}$} (done)
(reserved) edge[bend left=30] node[above,sloped,pos=0.6] {$\tup{cid,pid,c}$} (done.north)
;

\path[-]
(users) edge node[left,pos=0.4] {$\tup{uid}$} (login)
(products) edge node[below,pos=0.5] {$\tup{pid,n,c}$} (addprod)
(bonuses) edge node[below,pos=0.5] {$\tup{uid,bt'}$} (changebonus)
(bonuses) edge node[below,pos=0.5] {$\tup{uid,bt}$} (paybonus)
;
\path[X-stealth',thick,color=red]
(acqbonus) edge node[below] {$\tup{uid,cid,bt}$} (rebonus)
;

\end{tikzpicture}
}
\caption{The control of a \dbnet for online shopping. 
Here, $\nu_c$ is a fresh input variable corresponding to a newly created cart, whereas $dest$ is an arbitrary input variable representing a destination address. The rollback output arc (corresponds to the rollback flow) is in red and decorated with an ``x".} \label{fig:shopping-cart-cl}
\end{figure}
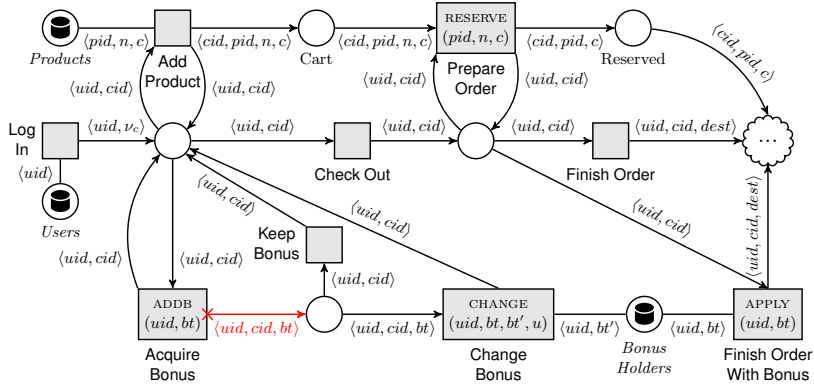

Figure~\ref{fig:shopping-cart-cl} shows the control layer of the shopping cart example. The queries specified in Example~\ref{ex:dl} are assigned to the corresponding view places: $\quass(\text{\emph{Products}}):=\cname{Q_{products}}$, $\quass(\text{\emph{Users}}):=\cname{Q_{users}}$ and 
$\quass(\text{\emph{Bonus Holders}}):=\cname{Q_{wbonus}}$. The actions (with their formal parameters) assigned to transitions via $\aass$ graphically appear in grey transition boxes. 

The execution semantics of a \dbnet simultaneously accounts for the progression of a database instance compliant with the persistence layer of the net, and for the evolution of a marking over the control layer of the net. Due to space limitations, we refer to the definition of the formal semantics studied in~\cite{MonR17}. 
We thus assume that the execution semantics of both \nucpns and \dbnets can be captured with a possibly infinite-state labeled transition system (LTS) that accounts for all possible executions starting from their initial markings. 
While transitions in such LTSs model the effect of firing nets under given bindings, their state representations slightly differ. Namely, in the case of \nucpns we have markings (like, for example, in coloured Petri nets~\cite{JeK09}), while in the case of \dbnets one also has to take into account database states. 
W.l.o.g., we shall use $\tsys{M_0}{\cpn} = \tup{\M,M_0,\trans,L}$ to specify an LTS for a \nucpn $\cpn$ with initial marking $M_0$ and $\tsys{\istate}{\dbn} = \tup{\states,\istate,\trans,L}$ to specify an LTS for a \dbnet $\dbn$ with initial snapshot $\istate=\tup{\I_0,m_0}$, where $\I_0$ is the initial database instance and $m_0$ is the initial marking of the control layer.

\section{Translation}
We are now ready to describe the translation from \dbnets to \nucpns with priorities (we assume the reader is familiar with transition priorities). Recall that this is not just of theoretical interest, but has also practical implications. In \cite{RMRS18}, we have presented a prototypical implementation of \dbnets in CPN Tools that, using Access/CPN and Comms/CPN, allow to model and simulate \dbnets. 
However, we realized that CPN Tools would not correctly generate the state space of the \dbnet at hand. This is due to the fact that the CPN Tools state space construction module does not consider third-party extensions which, in our setting, implies that the content of the view places is not properly recomputed after each transition firing. 

The first challenge to overcome is how the the database schema is represented in the target net. To this aim, we introduce special \emph{relation places} that copy corresponding database relations
by mirroring their signature to the type definitions of places.\footnote{Relation places do not differ from the normal \nucpn places. 
We use the different name in order to conceptually distinguish their origin.} 
In this light, database instances will correspond to relation place markings, where tokens are nothing but tuples. 
All other \dbnet elements (for example, bindings for fresh variables, action execution) require actual computation that happens 
when a transition fires.
Intuitively, every \dbnet transition \transition{T} is represented using the following four phases:
\begin{compactenum}[1.]
\item Collect bindings and compute the content of view places adjacent to \transition{T}.
\item If there is an action assigned to \transition{T}, execute it. We employ auxiliary boolean places that control 
whether an update has actually happened (that is, a token representing a tuple has been removed from or added to a relation place). 
\item Check the satisfaction of integrity constraints.
\item Finish the computation and generate a new marking.
\begin{compactenum}
	\item If all constraints are satisfied, empty the auxiliary boolean places used in 2), release the lock, and 
	populate the postset of \transition{T}.
	\item  If some constraint is violated, roll-back the effects. This is done in reverse order w.r.t. phase 2), applying or skipping a reverse
	 update depending on how the values in the special places. 
	 After this, the relation places have the content they had before the action was applied. Then, one releases the lock and pushses the special postset corresponding to the roll-back arc (if any) attached to \transition{T}.  
\end{compactenum}
\end{compactenum}
To realize the execution of an original \dbnet transition, all the four phases are executed uninterruptedly (under lock).
In the reminder of the section we formalizing the phases discussed above.

\begin{figure}[t!]
\centering
\resizebox{.72\hsize}{!}{
\begin{tikzpicture}[->,>=stealth',auto,x=1.3cm,y=1.0cm,thick]

   \node[transition,label={below:\transition{T}}] (t) at (3,0) {~~~~\large $\actname{Act}(\vec{x})$~~~~};
   	\node[label=\large $\mathbf{[}{\color{mgreen}G(\vec{y})}\mathbf{]}$]  (guard) at (1.8,0.07) {};

    \viewplace{vp1}{(1.6,1.8)}{left:\large $V_1$};
    \node[label=\small{$\ldots$}]  (dots1) at (3,1.6) {};
    \viewplace{vpm}{(4.4,1.8)}{left:\large $V_m$};
	
	\node[above right=0mm of vp1]  (vq1) 
	  {\large $\cname{Q}_{\cname{V_1}} (\vec{x}_1)$};
    \node[above right=0mm of vpm]  (vqm)
      {\large $\cname{Q}_{\cname{V_m}} (\vec{x}_m)$};

\node[
    placelem,
    label=above:{\large input places}
	](start) at (0,0) {\tiny{$\circ \circ \circ$}};
 
 \node[
    placelem,
    label=above:{\large output places}
	](end) at (6,.9) {\tiny{$\circ \circ \circ$}};
	
 \node[
    placelem,
    label=above:{\large rollback places}
	](rollback) at (8,0) {\tiny{$\circ \circ \circ$}};

  \path[]
    (start) edge node[below,xshift=-1mm] {\large $\vec{z}$} (t)
    ($(t.east)+(0,0.15)$) edge node[sloped,above] {\large $\vec{o}$} (end);
  \path[red,x-stealth']
   ($(t.east)$)edge node[sloped,below] {\large $\vec{r}$} (rollback)

    ;
    \path[-]
	(vp1) edge node[below,sloped,xshift=-4mm] {\large $\vec{x}_1$} (t)
	(vpm) edge node[above,sloped,xshift=0mm] {\large $\vec{x}_m$} (t)
	;
\end{tikzpicture}}
\caption{A generic \dbnet transition accessing multiple view places} 
\label{fig:dnet-generic}
\end{figure}
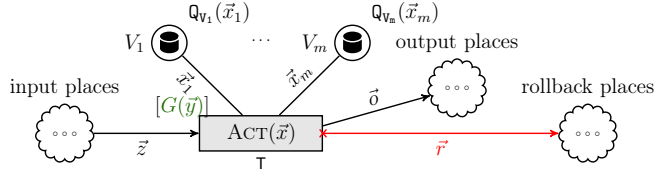
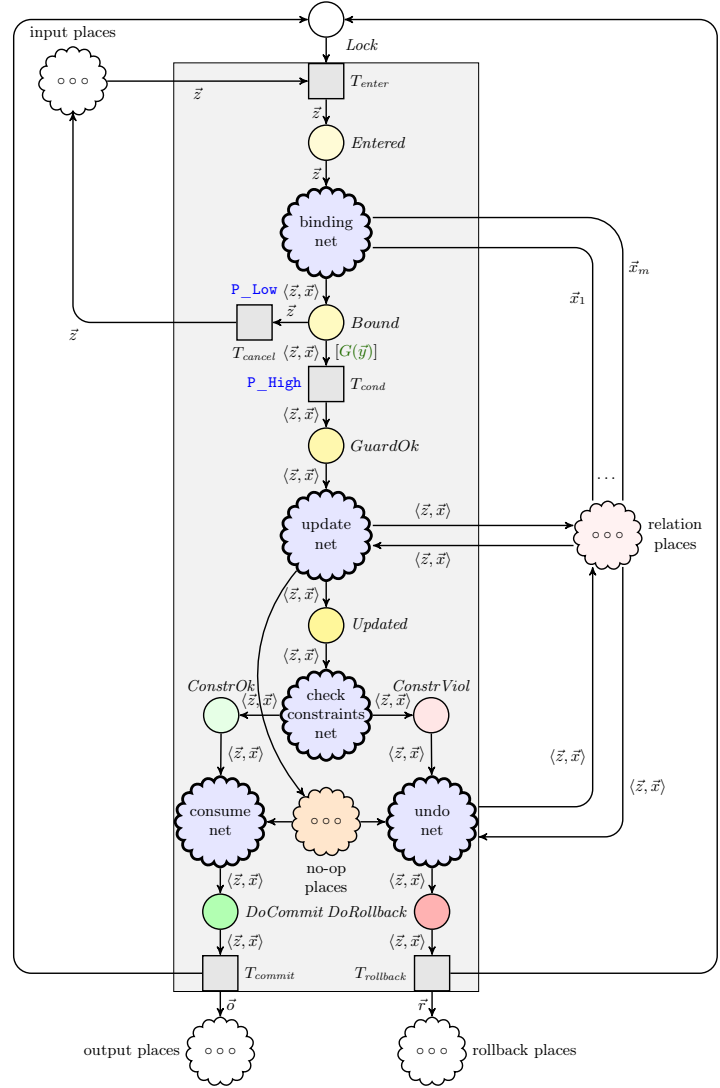
\begin{figure}[t!]
\centering
\resizebox{.78\textwidth}{!}{
  \begin{tikzpicture}
  [x=1cm,y=1cm,node distance=.5cm,thick,font=\footnotesize]
    
%

\node[
    placelem,
    label=above:input places
	](input) at (0,0) {\small{$\circ \circ \circ$}};
	
\node[
    transition,
    right=4cm of input,
    label=right:$T_{\mathit{enter}}$
    ] (tenter) {};

\node[
    place,
    above=of tenter,
    label=below right:$\mathit{Lock}$
    ] (plock) {};

\node[
    enteredplace,
    below=of tenter,
    label=right:$\mathit{Entered}$
    ] (pentered) {};

\node[
    netelem,
    below=of pentered,
    label=center:\begin{tabular}{@{}c@{}}binding\\net\end{tabular}
    ](binding) {};	


\node[
  boundplace,
  label=right:{$Bound$},
  below=of binding
  ] (pbound) {};

\node[
    transition,
    xshift=-2mm,
    left=of pbound,
    label=below:$T_{\mathit{cancel}}$
    ] (tcancel) {};
\node[above=0mm of tcancel] {\plow};

\node[
    transition,
    below=of pbound,
    label={right:$T_{\mathit{cond}}$},
    ] (tcond) {};
\node[left=0mm of tcond] 
  {\phigh};
\node[above=0mm of tcond,xshift=6mm] 
  {$\mathbf{[}{\color{mgreen}G(\vec{y})}\mathbf{]}$};  

\node[
  guardokplace,
  label=right:{$GuardOk$},
  below=of tcond
  ] (pguard) {};

\node[
  netelem,
  below=of pguard,
  label=center:\begin{tabular}{@{}c@{}}update\\net\end{tabular}
  ] (update) {};
  
\node[
  updatedplace,
  label=right:{$\mathit{Updated}$},
  below=of update
  ] (pupdated) {};

\node[
    relationselem,
    label=right:{\begin{tabular}{@{}c@{}}relation\\places\end{tabular}},
    right=4cm of update
	](relations) {\small{$\circ \circ \circ$}};
\node[
  above=3mm of relations] {\ldots};

\node[
  netelem,
  below=of pupdated,
  label=center:\begin{tabular}{@{}c@{}}check\\constraints\\net\end{tabular}
  ] (check) {};    

\node[
    noopelem,
    label=below:{\begin{tabular}{@{}c@{}}no-op\\places\end{tabular}},
    below=of check
	](noop) {\small{$\circ \circ \circ$}};

\node[
  constrokplace,
  label=above:{$\mathit{ConstrOk}$},
  left=of check,
  xshift=-3mm
  ] (pcok) {};
  
\node[
  violplace,
  label=above:{$\mathit{ConstrViol}$},
  xshift=3mm,
  right=of check
  ] (pcviol) {};

\node[
  netelem,
  ultra thick,
  left=of noop,
  label=center:\begin{tabular}{@{}c@{}}consume\\net\end{tabular}
  ] (consume) {};    

\node[
  netelem,
  ultra thick,
  right=of noop,
  label=center:\begin{tabular}{@{}c@{}}undo\\net\end{tabular}
  ] (undo) {};    

\node[
  dorollbackplace,
  label=left:$\mathit{DoRollback}$,
  below=of undo
  ] (prb) {};

\node[
  docommitplace,
  label=right:$\mathit{DoCommit}$,
  below=of consume
  ] (pc) {};
  
\node[
  transition,
  label=left:$T_\mathit{rollback}$,
  below=of prb
  ] (trb) {};
  
\node[
  transition,
  label=right:$T_\mathit{commit}$,
  below=of pc
  ] (tc) {};
  
\node[
    placelem,
    label=left:{output places},
    below=of tc
	](output) {\small{$\circ \circ \circ$}};

 \node[
    placelem,
    label=right:{rollback places},
    below=of trb
	](rollback) {\small{$\circ \circ \circ$}};

\path[-stealth',thick]
  (input) 
    edge node[below left] {$\vec{z}$}  
  (tenter)
  (plock)
    edge
  (tenter)
  (tenter) 
    edge node[left] {$\vec{z}$}  
  (pentered)
  (pentered) 
    edge node[left] {$\vec{z}$}  
  (binding)
  (binding) 
    edge node[left] {$\tup{\vec{z},\vec{x}}$}
  (pbound)
  (pbound)
    edge node[above] {$\vec{z}$}
  (tcancel)
  (pbound) 
    edge node[left] {$\tup{\vec{z},\vec{x}}$}
  (tcond)
  (tcond) 
    edge node[left] {$\tup{\vec{z},\vec{x}}$}
  (pguard)
  (pguard)
    edge node[left] {$\tup{\vec{z},\vec{x}}$}  
  (update)
  (update)
    edge[out=-130,in=130]
  (noop)
  ($(update.east)+(0,2mm)$)
    edge node[above,pos=0.3] {$\tup{\vec{z},\vec{x}}$}
  ($(relations.west)+(0,2mm)$)
  ($(relations.west)-(0,2mm)$)
    edge node[below,pos=0.7] {$\tup{\vec{z},\vec{x}}$}
  ($(update.east)-(0,2mm)$)
  (update)
    edge node[left] {$\tup{\vec{z},\vec{x}}$}
  (pupdated)
  (pupdated)
    edge node[left] {$\tup{\vec{z},\vec{x}}$}
  (check)
  (check)
    edge node[above] {$\tup{\vec{z},\vec{x}}$}
  (pcok)
  (check)
    edge node[above] {$\tup{\vec{z},\vec{x}}$}
  (pcviol)
  (pcok)
    edge node[right] {$\tup{\vec{z},\vec{x}}$}
  (consume)
  (pcviol)
    edge node[left] {$\tup{\vec{z},\vec{x}}$}
  (undo)
  (noop) edge (undo)
  (noop) edge (consume)
  (consume)
      edge node[right] {$\tup{\vec{z},\vec{x}}$}
  (pc)
  (undo)
      edge node[left] {$\tup{\vec{z},\vec{x}}$}
  (prb)
  (pc)
      edge node[right] {$\tup{\vec{z},\vec{x}}$}
  (tc)
  (prb)
      edge node[left] {$\tup{\vec{z},\vec{x}}$}
  (trb)
  (tc)
    edge node[right] {$\vec{o}$}
  (output)
  (trb)
    edge node[left] {$\vec{r}$}
  (rollback)
  ;
     
\draw[-,rounded corners=10pt,thick]
  ($(binding.east)-(0,3mm)$)
    -| node[left,yshift=-1cm] {$\vec{x}_1$} 
  ($(relations.north)-(3mm,0)$);
\draw[-,rounded corners=20pt,thick]
  ($(binding.east)+(0,3mm)$)
    -| node[right,yshift=-1cm] {$\vec{x}_m$} 
  ($(relations.north)+(3mm,0)$);

\draw[-stealth',rounded corners=10pt,thick]
  (tcancel)
    -| node[below] {$\vec{z}$}
  (input);
\draw[-stealth',rounded corners=10pt,thick]
   ($(relations.south)+(3mm,0)$)
    |- node[right,yshift=1cm] {$\tup{\vec{z},\vec{x}}$} 
  ($(undo.east)-(0,3mm)$);
\draw[-stealth',rounded corners=10pt,thick]
   ($(undo.east)+(0,3mm)$)
    -| node[left,yshift=1cm] {$\tup{\vec{z},\vec{x}}$} 
  ($(relations.south)-(3mm,0)$);
\draw[-stealth',rounded corners=10pt,thick]
  (trb)
  -|  
  ($(relations.east)+(15mm,0)$)
  |-
  (plock);
\draw[-stealth',rounded corners=10pt,thick]
  (tc)
  -|  
  ($(input.west)-(5mm,0)$)
  |-
  (plock);

\begin{pgfonlayer}{background}
  \node[
    draw,
    rectangle,
    fill=gray!10,
    fit={(tc) (trb) (consume) (undo) (tenter)},
    inner sep=0,
  ] (t) {};
\end{pgfonlayer}
\end{tikzpicture}
}
\caption{Overall \nucpn encoding of the \dbnet transition shown in Figure~\ref{fig:dnet-generic}. Blue clouds stand for subnets that are expanded next, and $\vec{x}$ is a shortcut for the tuple consisting of $\vec{x}_1\,\ldots,\vec{x}_m$. Elements within the gray rectangle are local to the transition, whereas external elements are shared at the level of the whole net.}
\label{fig:cpn-encoding}
\end{figure}

A generic \dbnet $\dbn_\tau$ that we use to demonstrate the translation is represented in Figure~\ref{fig:dnet-generic}. 
Here, we assume that \transition{T} is contains enough of tokens assigned by its input flow and its eventual firing is subject to the $G(\vec{y})$ guard evaluation. 
$\vec{y}$, in turn, is bound to values form $\vec{z}$ and from $m\in\naturals$ ordered view places, where each view place $V_i$ has a query $\cname{Q}_\cname{V_i}$ assigned to it. 
The \nucpn $\cpn_\tau$ representing $\dbn_\tau$ is depicted in Figure~\ref{fig:cpn-encoding}.
To facilitate the translation, we make three working hypothesis. First, we assume that the relational schema is equipped only with three types of constraints: primary keys, foreign keys and domain constraints. 
Second, for ease of presentation, we assume that the resulting \nucpn model can deal with \dbnets external variables.
This assumption, however, is correct from the practical point of view as it has been already shown before that a preliminary implementation of \dbnets in CPN Tools ~\cite{RMRS18}
provides functionality necessary for computing bindings for external variables.  Third, we naturally extend the notion of \nucpn with read arcs.

\subsection{Computing Views Using CPN Places}
\label{sec:encoding-views}
\begin{figure}[t!]
\centering
\resizebox{.90\hsize}{!}{
\begin{tikzpicture}
\footnotesize
\begin{scope}[->,>=stealth',auto,x=2.2cm,y=1.7 cm,thick]
  \viewplace{products}{(0,1.3)}{below,xshift=.0cm:\emph{Products}};
  \node[place] (logged) at (1,0.3) {};
  \node[transition, label=below:$\begin{array}{@{}c@{}} \text{\transition{Add}} \\ \text{\transition{Product}}\end{array}$] (addprod) at (1,1.3) {};
	\node[place,label=below:Cart] (cart) at (2.35,1.3) {};

\path
(logged) edge[bend left=50] node[left,pos=0.2] {$\tup{uid,cid}$}  (addprod.south west)
(addprod.south east) edge[bend left=50] node[right,pos=0.8] {$\tup{uid,cid}$}  (logged)
(addprod) edge node[below] {$\tup{cid,pid,n,c}$}  (cart)
;

\path[-]
(products) edge node[below] {$\tup{pid,n,c}$} (addprod)
;
\end{scope}


\begin{scope}[yshift =2.7cm,xshift=7.5cm,relation/.style={rectangle split, rectangle split parts=#1, rectangle split part align=base, draw, anchor=center, align=center, text height=3mm, text centered}]\hspace*{-0.3cm}


\node (Product_title) {\relname{Product}};
\node [relation=1, rectangle split horizontal, rectangle split part fill={lightgray!50}, anchor=north west, below=0.6cm of Product_title.west, anchor=west] (Product)
{\underline{\attrname{Name}} : $\typename{string}$};

\node [below=1.5cm of Product_title.west, anchor=west] (Warehouse_title) {\relname{InWarehouse}};
\node [relation=3, rectangle split horizontal, rectangle split part fill={lightgray!50}, anchor=north west, below=0.6cm of Warehouse_title.west, anchor=west] (Warehouse)
{\underline{\attrname{PID}} :$\typename{int}$%
\nodepart{two} \attrname{name} : $\typename{string}$
\nodepart{three}  \attrname{cost} : $\typename{real}$};


\draw[-latex] ($(Warehouse.two north) +(0.3,0)$) |- node[rectangle,draw=black,fill=white,xshift=13mm,yshift=-6mm,inner sep=.5mm,minimum size=1mm]{\footnotesize{\textup{FK\_\relname{InWarehouse}\_\relname{Product}}}} ($(Product.one east) + (0,0)$);

\end{scope}
\end{tikzpicture}
}
\caption{Fragments of the process and persistence layers of  the \dbnet in Figure~\ref{fig:shopping-cart-cl}} 
\label{fig:dbnets-view-trick-1}
\end{figure}
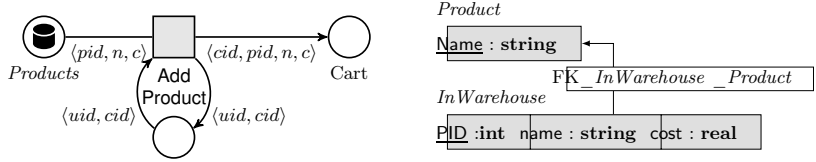

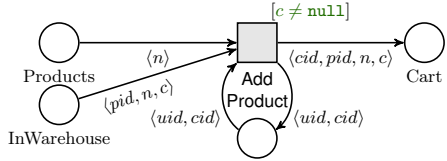
\begin{figure}[t!]
\centering
\resizebox{.50\hsize}{!}{
\begin{tikzpicture}[->,>=stealth',auto,x=2.2cm,y=1.7 cm,thick]
  \node[place,label=below:Products] (products) at (-.6,1.3) {};
   \node[place,label=below:InWarehouse] (warehouse) at (-.6,0.65) {};
  \node[place] (logged) at (1,0.3) {};
  \node[transition, label=below:$\begin{array}{@{}c@{}} \text{\transition{Add}} \\ \text{\transition{Product}}\end{array}$] (addprod) at (1,1.3) {};
  \node (guard) at (1.43,1.63) {$\mathbf{[}{\color{mgreen}c\neq\cname{null} }\mathbf{]}$};
	\node[place,label=below:Cart] (cart) at (2.35,1.3) {};

\path
(logged) edge[bend left=50] node[left,pos=0.2] {$\tup{uid,cid}$}  (addprod.south west)
(addprod.south east) edge[bend left=50] node[right,pos=0.8] {$\tup{uid,cid}$}  (logged)
(addprod) edge node[below] {$\tup{cid,pid,n,c}$}  (cart)
(products) edge node[below] {$\tup{n}$} (addprod)
(warehouse) edge node[below,sloped,pos=0.35] {$\tup{pid,n,c}$} (addprod)

;
\end{tikzpicture}
}
\caption{A \nucpn representation of the \dbnet in Figure~\ref{fig:dbnets-view-trick-2}} 
\label{fig:dbnets-view-trick-2}
\end{figure}

We start by describing how the view computation should work using only \nucpn places. 
Let us consider as an example a subnet $\dbn_{tr}$ of the \dbnet present in Figure~\ref{fig:shopping-cart-cl} that models only the selection of available products. To access products that are available in the warehouse and that have prices assigned to them, we need to run a query $\cname{Q}_{\cname{products}}(pid,n,c)=\relname{Product}(n)\land\relname{InWarehouse}(pid,n,c)\land c\neq\cname{null}$. 
Interestingly, such a query can be formulated directly using standard elements of \nucpns. Indeed, we may transfer the \dbnet in Figure~\ref{fig:dbnets-view-trick-1} into a \nucpn $\cpn_{tr}$ in Figure~\ref{fig:dbnets-view-trick-2} representing the project selection step. As one can see, the relations of $\dbn_{tr}$ have been copied to the same-named 
relation places, when $\cname{Q}_{\cname{products}}$ is treated as follows: 
$\cpn_{tr}$ accesses relation places with read-arcs (that have relation attributes as their inscriptions) so as to realize the projection, while the filter (i.e., $c\neq\cname{null}$) is basically plugged into the guard of \transition{Add Product}.
The result of the query is then propagated into the post-set of \transition{Add Product} using the free variables of $\cname{Q}_{\cname{products}}$ (i.e., $pid$, $n$ and $c$) in the arc inscriptions.

However, one may see that not every query can be handled when only using standard \nucpn elements. Assume a query $\cname{Q_{\neg available}}(n)=\relname{Product}(n)\land\neq\exists pid,c.\relname{InWarehouse}(pid,n,c)$ that lists products not available in the warehouse. In order to represent $\cname{Q_{\neg available}}$ in a \nucpn, one would need to extend the net with constructs allowing to fire a transition only if a certain element does not exists in a place incident to it. 
Thus, we restrict ourselves to the union of conjunctive queries with negative filters (or atomic negations) ($\text{UCQFs}^{\neq}$), that is $\fod{\types}$ queries of the form $\bigvee_{i=1}^n \exists\vec{y_i}.conj_i(\vec{x})$, where $conj_i(\vec{x})$ is also a $\fod{\types}$ query that is a conjunction of relations $R(\vec{z})$, predicates $P(\vec{y})$ and their negations $\neg P(\vec{y})$. Henceforth, we use $\queries^{UCQF^{\neq}}$ to define a $\text{UCQF}^{\neq}$ subset of $\queries$. In SQL, a conjunctive query is a query representable with a SELECT-FROM-WHERE expression. 
As it has been already shown, the filter conditions (of the $\text{UCQFs}^{\neq}$ attached to view places) can be modeled using transition guards.

\begin{figure}[t!]
\centering
\resizebox{\textwidth}{!}{
\begin{tikzpicture}
  [x=1cm,y=1cm,node distance=2.5cm,thick,font=\footnotesize]
  
\node[
  enteredplace,
  label=below:$\mathit{Entered}$
  ] (pentered) {};
  
\node[
  transition,
  label=below:$\mathit{ComputeV_1}$,
  right=of pentered
  ] (tv1) {};
\node[
  above left=0mm of tv1
  ] {$\mathbf{[}{\color{mgreen}F_{V_1}(\vec{y}_1)}\mathbf{]}$};

\node[
  place,
  label=below:$\mathit{{V_{1}}Computed}$,
  right=of tv1
  ] (pv1) {};
  
\node[right=of pv1] (dots) {...};

\node[
  transition,
  label=below:$\mathit{ComputeV_m}$,
  right=of dots,
  xshift=5mm,
  ] (tvm) {};
\node[
  above right=0mm of tvm
  ] {$\mathbf{[}{\color{mgreen}F_{V_m}(\vec{y}_m)}\mathbf{]}$};

\node[
  boundplace,
  label=below:$\mathit{Bound}$,
  right=of tvm
  ] (pbound) {};
  
\node[
  relationselem,
  label=below left:relation places,
  above=1cm of dots
  ] (relations) {$\circ \circ \circ$};
  
\path[-stealth',thick]
(pentered)
  edge node[below] {$\vec{z}$}
(tv1)
(tv1)
  edge node[below] {$\tup{\vec{z},\vec{x}_1}$}
(pv1)
(pv1)
  edge node[below] {$\tup{\vec{z},\vec{x}_1}$}
(dots)
(dots)
  edge node[below] {$\tup{\vec{z},\vec{x}_1,\ldots,\vec{x}_{m-1}}$}
(tvm) 
(tvm)  
  edge node[below] {$\tup{\vec{z},\vec{x}}$}
(pbound);  

\draw[-,rounded corners=20pt,thick]
($(tv1.north)-(2mm,0)$)
  |- node[below left] {$\vec{x}_{K_1^1}$}
($(relations.west)+(0,2mm)$)
;
\draw[-,rounded corners=10pt,thick]
($(tv1.north)+(2mm,0)$)
  |- node[below right] {$\vec{x}_{K_1^{n_1}}$}
($(relations.west)-(0,2mm)$)
;
\node[above=2mm of tv1,rotate=90,anchor=west] {\ldots};

\draw[-,rounded corners=10pt,thick]
($(tvm.north)-(2mm,0)$)
  |- node[below left] {$\vec{x}_{K_m^1}$}
($(relations.east)-(0,2mm)$)
;
\draw[-,rounded corners=20pt,thick]
($(tvm.north)+(2mm,0)$)
  |- node[below right] {$\vec{x}_{K_m^{n_m}}$}
($(relations.east)+(0,2mm)$)
;
\node[above=2mm of tvm,rotate=90,anchor=west] {\ldots};

\path[-,dashed,thick]
($(dots.north)-(2mm,0)$) edge ($(relations.south)-(2mm,1mm)$); 
\path[-,dashed,thick]
($(dots.north)+(2mm,0)$) edge ($(relations.south)+(2mm,-1mm)$); 
\node[above=2mm of dots,rotate=90,anchor=west] {\ldots};

\end{tikzpicture}  
}
\caption{Expansion of the binding net from Figure~\ref{fig:cpn-encoding}}
\label{fig:phase1}
\end{figure}
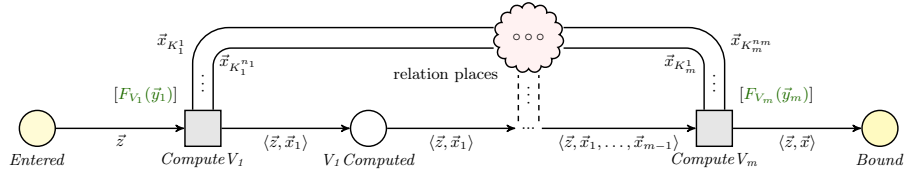

In case of multiple view places attached to one transition, we construct a net that computes them in a sequential manner. 
One may see the computation process as a pipeline. Whenever a transition that corresponds to a certain view place is enabled,
it fires and generated tokens that represent one of the tuples of the view. Then, acquired tokens are transferred to the next transition using variables in the arc inscriptions. The computation continues until the last view.                                                                                                                                                                                                                                                                                                                     
After that, the results of all the computations are transferred to the corresponding places, following the topology (i.e., the organization of arcs defined by the flow relations) of the original \dbnet. Note that the order in which views are computed has to be 
the same as the one defined for $\dbn_\tau$. 

A \nucpn in Figure~\ref{fig:phase1} shows how bindings and view places are computed in the case of the generic \dbnet $\dbn_\tau$. 
The computation process per view $V_i$ is realized by a transition called $\mathit{ComputeV_i}$ and analogous to the one explained before: we read necessary data from relation places,  representing
relations used in $\cname{Q}_\cname{V_i}$, and filter these data by means of $F_{V_i}(\vec{y})$. Note that variables on every read-arc 
adjacent to $\mathit{ComputeV_i}$ represent attributes of some relation $R$. 
The intermediate result of the view computation is then stored in a place called $\mathit{V_iComputed}$. 
As one can see from Figure~\ref{fig:phase1}, all the intermediate results are accumulated along the computation cycle. 
Moreover, we carry data provided with input variables of \transition{T} 
so as to check the validity of the guard $G$ (see Figure~\ref{fig:cpn-encoding}). 
This is done using prioritized transition $T_{cond}$. If the guard is not satisfied, one has to reset the computation process
by returning tokens that have been consumed at the beginning of the view computation (that is,  tokens that have been assigned to $z$).
We resolve this issue by introducing an auxiliary transition called $T_{cancel}$ that may fire only when the guard has been evaluated to $\false$. 
Scheduling between $T_{cond}$ and $T_{cancel}$ is managed 
by means of two priority labels \phigh and \plow (where $\phigh > \plow$) respectively assigned to them.

\subsection{Modeling RDBMS updates in CPNs}
\label{sec:encoding-updates}
We now show how database updates exploited by \dbnets could 
be represented using regular coloured Petri nets. We recall, that actions assigned to 
\dbnet transitions support addition and deletion of $\schema$-fact, 
which should preserve the set semantics adopted by the persistence layer. 

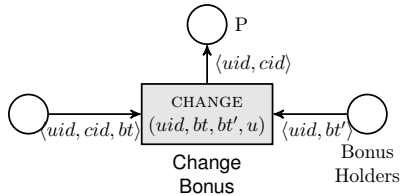
\begin{wrapfigure}[7]{l}{.5\textwidth}
\vspace*{-.2cm}
\resizebox{.45\textwidth}{!}{
\begin{tikzpicture}[->,>=stealth',auto,x=2.2cm,y=0.8cm,thick,scale=0.9,every node/.style={transform shape}]

 \node[place, label=below:$\begin{array}{@{}c@{}} \text{{Bonus}} \\ \text{{Holders}}\end{array}$] (bonuses) at (5.2,-2) {};

 \node[place,label=right:P] (logged) at (3.9,0) {};

	\node[place] (rebonus) at (2.45,-2) {};
	
	 \node[transition, label=below:$\begin{array}{@{}c@{}} \text{\transition{Change}} \\ \text{\transition{Bonus}}\end{array}$] (changebonus) at (3.9,-2) 
  {$\begin{array}{@{}c@{}}
                 \text{\actname{change}}\\
  				(uid,bt,bt',u)
               \end{array}
               $ };

\path
(rebonus) edge node[below,pos=0.45] {$\tup{uid,cid,bt}$} (changebonus)
(changebonus.north) edge node[right,pos=0.5] {$\tup{uid,cid}$} (logged)
(bonuses) edge node[below,pos=0.4] {$\tup{uid,bt'}$} (changebonus)
;
\end{tikzpicture}
}
\caption{A subnet of the \dbnet in Figure~\ref{fig:shopping-cart-cl} describing the bonus change step} 
\label{fig:dbnets-add-trick-1}
\end{wrapfigure}

In Figure~\ref{fig:dbnets-add-trick-1} we consider a \dbnet describing the bonus change step of he online shopping process. 
Here, for ease of presentation, instead of considering a view place for bonus holders, we use a regular (control) place that stores the same kind of data. 

The translation of \dbnet-like database updates into \nucpns is conceptually similar to the representation of the view computation process: 
\dbnet actions must be performed sequentially within a critical section that can be entered whenever a special write lock is available. 
For preserving the set semantics over every relation place, we use prioritized  transitions so as to check whether a tuple to be added or deleted already exists in the relation place. 
Specifically, for each tuple we would introduce two transitions, one with a higher priority and another with a lower priority, and an auxiliary (\emph{no-op}) boolean place.
The first transition can fire if the tuple is in the corresponding relation place, while the second one would fire otherwise. 
Both transitions are adopted to deal with additions and deletions. 
In case of additions, the highly 
prioritized transition would not add the tuple, while the
one with the lower priority would do otherwise. 
To deal with deletions, we mirror the previous case: if the tuple exists, 
then one can safely remove it; otherwise, one proceeds without changes.  
Upon firing of any of these transitions, the auxiliary place receives a boolean token. 
If the value of the token is $\true$, then it means that the tuple has been successfully added or deleted. 
In case the database update has not taken place,
the token value is going to be $\false$. 
It is important to note that the update execution order of \dbnet actions must be also preserved in their \nucpn representation. 
That is, for every action $\action$ we first delete all the tuples from $\adel{\action}$, and only then add those from $\aadd{\action}$.

\begin{figure}[t!]
\hspace{-0.5cm}
\begin{tikzpicture}[->,>=stealth', auto,x=1.4cm,y=1.2cm,thick,scale=0.75,every node/.style={transform shape}]
 
  \node[place] (ready) at (-0.2,0) {};
  
  \node[place,label=left:Bonus Holders]  (bonuses) at (1.8,1.5) {};

   \node[transition,label={below:$\begin{array}{@{}c@{}}
                 \text{\transition{Change}}\\
  				\text{\transition{Bonus}}
               \end{array}
               $}] (change) at (1.8,0) {~~~~~~~~~~};
               
  \node[place,label=left:\placename{WithBonus}] (wbonus) at (4.5,2.6) {};

  \node[place,label=below:$D1$] (del1) at (3,0) {};
  \node[place,label={[xshift=-1mm]right:\placename{DoneD1}}] (dd1) at (4.5,0) {};
 \node[transition,label={right:\transition{ExistsD1}}] (exists1) at (4.5,1.5) {~~~~};
 \node[label=\phigh]  (hpriority_del1) at (3.8,1.5) {};
  \node[transition,label={right:\transition{NotExistsD1}}] (notexists1) at (4.5,-1.5) {~~~~};
 \node[label=\plow]  (lpriority_del1) at (3.9,-1.95) {};

  \node[place,label=below:$A1$] (add1) at (6,0) {};
      \node[place,label={[xshift=-1mm]right:\placename{DoneA1}}] (da1) at (7.5,0) {};
\node[transition,label={right:\transition{NotExistsA1}}] (notexists3) at (7.5,1.5) {~~~~};     
 \node[label=\plow]  (lpriority_add1) at (6.8,1.50) {};          
  \node[transition,label=right:\transition{ExistsA1}] (exists3) at (7.5,-1.5) {~~~~};
 \node[label=\phigh]  (hpriority_add1) at (6.8,-1.95) {};

   \node[place,label=right:P]  (logged) at (9.5,0) {};
     
	\node[place,label=right:\placename{Lock}]  (lock) at (11,0) {};

  \path[]
    (ready) edge node[below,pos=0.45] {$\tup{uid,cid,bt}$} (change)
    (bonuses) edge node[left,xshift=-.5mm] {$\tup{uid,bt'}$} (change)
    
	(change) edge node[below,xshift=-1mm] {$\xi$} (del1.west)
	(del1) edge node[left,xshift=-0.2mm,yshift=1mm] {$\xi$} (exists1)
	(del1) edge node[left,xshift=-0.2mm,yshift=-1mm] {$\xi$} (notexists1)
	(exists1) edge node[right,xshift=-0.2mm,yshift=-1mm] {$\tup{\true}$} (dd1)
	(notexists1) edge node[right,xshift=-0.2mm,yshift=1mm] {$\tup{\false}$} (dd1)
	(wbonus) edge node[left,xshift=-0.2mm,yshift=1mm] {$\tup{uid,bt}$} (exists1)
	(exists1.south east) edge node[sloped,right,xshift=-7mm,yshift=2.5mm] {$\xi$} (add1.north west)
	(notexists1.north east) edge node[sloped,right,xshift=-7mm,yshift=-2.5mm] {$\xi$} (add1.south west)
	(add1) edge node[left,xshift=-0mm,yshift=0mm] {$\xi$} (exists3)
	(add1) edge node[left,xshift=-2mm,yshift=-2mm] {$\xi$} (notexists3)
		(exists3) edge node[right,xshift=-0.2mm,yshift=1mm] {$\tup{\false}$} (da1)
	(notexists3) edge node[right,xshift=-0.2mm,yshift=-1mm] {$\tup{\true}$} (da1)
	(notexists3.north) edge[bend right=25] node[rotate=-10,sloped,right,xshift=0mm,yshift=3mm,pos=0.4] {$\tup{uid,bt'}$} (wbonus.east)
	(notexists3.south east) edge node[sloped, right,xshift=-9.2mm,yshift=2.5mm] {$\tup{sid,tid}$} (logged.north west)
	(exists3.north east) edge node[sloped,right,xshift=-9.2mm,yshift=-2.5mm] {$\tup{sid,tid}$} (logged.south west)
	(lock.south) edge[bend left=50] node[above] {$\tup{e}$} ($(change.south east)+(0,-0.1)$)
	(notexists3.north east) edge[bend left=30] node[above] {$\tup{e}$} (lock.north west)
	(exists3.south east) edge[bend right=30] node[above] {$\tup{e}$} (lock.south west)
    ;
    
     \draw[-] ($(exists3.north)+(-0.1,0)$) to[out=-260, in=-25] node[rotate=-35,above,xshift=-15mm,yshift=2mm]{$\tup{uid,bt'}$}  (wbonus.south east);

\end{tikzpicture}
\vspace*{-1cm}
\caption{The CPN representation of the \dbnet in Figure~\ref{fig:dbnets-add-trick-1}, where $\xi$, for ease of reading, denotes the tuple $\tup{uid,cid,bt,bt'}$}
\label{fig:dbnets-add-trick-2}
\end{figure}
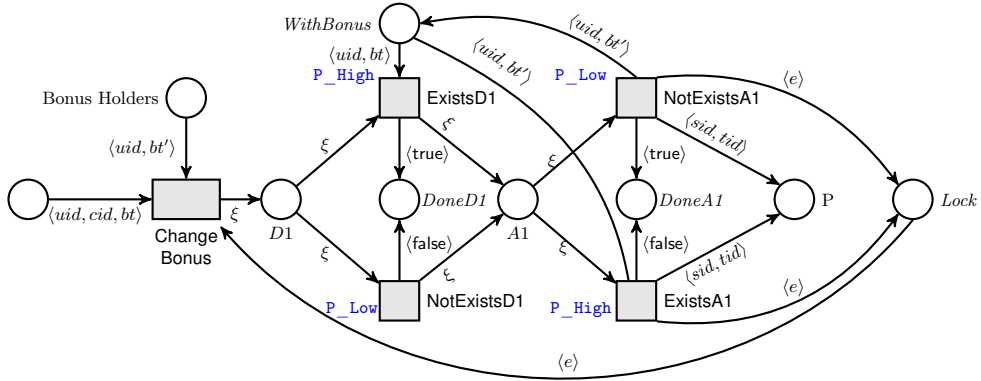

We incorporate aforementioned modeling guidelines in the \nucpn depicted in Figure~\ref{fig:dbnets-add-trick-2}.
Since $\actname{change}$ in $\dbn_{tr}^\alpha$ contains multiple database updates, the model starts with deleting $\relname{WithBonus}(uid,bt)$ from $\relname{WithBonus}$. To do so, at first one checks whether the relation place \placename{WithBonus} contains the tuple we would like to remove. 
This is done using \transition{Exists D1} that performs conditional removal of $\relname{WithBonus}(uid,bt)$, that is, if there is a token in \relname{WithBonus} such that bindings of inscriptions on $(D1,\transition{Exists D1})$ and $(\placename{WithBonus},\transition{ExistsD1})$ coincide, then \transition{ExistsD1} is enabled, and upon firing consumes the selected token from \placename{WithBonus} and populates one token with value $\tup{true}$ (the value $\true$ means that the update has been successfully accomplished) in \placename{DoneD1}. 
Note that \transition{Exists D1} is always checked first given the higher priority label assigned to it. If the tuple does not exist, then one proceeds with firing \transition{NotExistsD1} and populating one token with value $\false$ in \placename{DoneD1}.  
Now, when we reach the first control place allowing to perform
the add operation over \relname{WithBonus}, we start by checking whether \placename{WithBonus} already contains the $\relname{WithBonus}(uid,bt')$ tuple.
Specifically, we use the read arc $(\transition{ExistsA1},\placename{WithBonus})$ that has the only purpose of checking whether the token is present in the place. 
In case there is no token that matches values assigned to $\xi$, we proceed with adding $\relname{WithBonus}(pid,bt')$ with \transition{NotExistsA1} that has the lower priority label assigned to it and consequently populate a $\tup{\true}$-valued token in \placename{DoneA1}. 
Note that the whole computation process is ``guarded'' with the global lock variable (needed for the consequent execution of all the steps defined in Figure~{fig:cpn-encoding}): whenever started, the token is removed from it and can be returned only after the last operation of the action has been carried out.

\begin{figure}[t!]
\centering
\resizebox{\textwidth}{!}{
\begin{tikzpicture}
  [x=1cm,y=1cm,node distance=.9cm,thick,font=\footnotesize]
  
\node[
  guardokplace,
  label=below:$\mathit{GuardOk}$
  ] (pguard) {};

\node[
  netelem,
  label=center:
      \begin{tabular}{@{}c@{}}del net\\($R_{d,1}(\vec{y}_{d,1})$)\end{tabular},
  right=of pguard,
    minimum width=2cm,
  fill=red!10,
  ] (del1) {};
  
\node[
  place,
  right=of del1
  ] (pdel1) {};

\node[
  right=of pdel1
  ] (dotdel) {...};
  
 \node[
  place,
  right=of dotdel
  ] (pdeln) {}; 
  
\node[
  netelem,
  label=center:
      \begin{tabular}{@{}c@{}}add net\\($R_{a,1}(\vec{y}_{a,1})$)\end{tabular},
  right=of pdeln,
  fill=mgreen!10,
  minimum width=2cm,
  ] (add1) {};
  
\node[
  place,
  right=of add1
  ] (padd1) {};

\node[
  right=of padd1
  ] (dotadd) {...};
  
\node[
  updatedplace,
  label=below:$\mathit{Updated}$,
  right=of dotadd
  ] (pupdated) {};

\node[
  relationselem,
  label={[yshift=-3mm,xshift=1mm]left:relation places},
  above=of pdeln,
  yshift=-6mm
  ] (relations) {$\circ \circ \circ$};
  
\node[
  noopelem,
  label={[yshift=3mm,xshift=1mm]left:no-op places},
  yshift=7mm,
  below=of pdeln
  ] (noop) {$\circ \circ \circ$};

\path[-stealth']
(pguard)
  edge node[above] {$\tup{\vec{z},\vec{x}}$}   
(del1)
(del1)
  edge node[above] {$\tup{\vec{z},\vec{x}}$}   
(pdel1)
(pdel1)
  edge node[above] {$\tup{\vec{z},\vec{x}}$}   
(dotdel)
(dotdel)
  edge node[above] {$\tup{\vec{z},\vec{x}}$}   
(pdeln)
(pdeln)
  edge node[above] {$\tup{\vec{z},\vec{x}}$}   
(add1)
(add1)
  edge node[above] {$\tup{\vec{z},\vec{x}}$}   
(padd1)
(padd1)
  edge node[above] {$\tup{\vec{z},\vec{x}}$}   
(dotadd)
(dotadd)
  edge node[above] {$\tup{\vec{z},\vec{x}}$}   
(pupdated)
;

\draw[-stealth',rounded corners=10pt]
($(relations.west)+(0,2mm)$)
  -| node[left,yshift=-3mm] {$\vec{y}_{d,1}$}
($(del1.north)+(0,0mm)$);
\draw[-stealth',rounded corners=10pt]
(del1.south)
  |- node[right,yshift=2mm,xshift=3mm] {$\tup{b}$}
(noop.west);
\draw[-stealth',rounded corners=20pt]
($(add1.north)+(3mm,0)$)
  |- node[right,yshift=-5mm,xshift=-.5mm] {$\vec{y}_{a,1}$}
($(relations.east)+(0,2mm)$);

\draw[-stealth',rounded corners=10pt]
(add1.south)
  |- node[left,yshift=2mm,xshift=-3mm] {$\tup{b}$}
(noop.east)
;

\draw[-,rounded corners=10pt]
($(relations.east)-(0,1mm)$)
  -| node[left,yshift=-2.7mm,xshift=-2mm] {$\vec{y}_{a,1}$}
($(add1.north)+(-2mm,0)$);

\end{tikzpicture}  
}
\caption{Expansion of the update net from Figure~\ref{fig:cpn-encoding}}
\label{fig:update-net}
\end{figure}
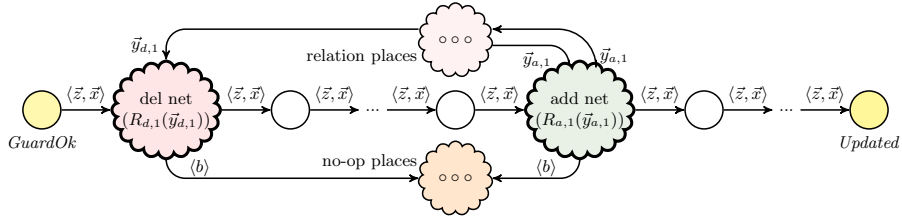

Next we show how an action is encoded considering the general \dbnet $\dbn$ in Figure~\ref{fig:dnet-generic}. Note that \transition{T} 
is equipped with an action  \actname{Act}, where some of the action parameters $\vec{x}$ coincide with external variables.
\actname{Act} is defined on top of $\pers$ with
$\apar{\actname{Act}} = \tup{x_1,\ldots,x_n}$,
	$\adel{\actname{Act}} = F^{-}$ and
	$\aadd{\actname{Act}} = F^{+}$,
where $F^{-}$ and $F^{+}$ are two sets of $\schema$-facts that should be respectively deleted and added. 
%
The CPN representing that expansion of the update net from Figure~\ref{fig:cpn-encoding} is depicted in Figure~\ref{fig:update-net}. 
The computation starts by checking the guard of \transition{T} with transition $T_{cond}$ (cf. Figure~\ref{fig:cpn-encoding}). 
If the guard evaluates to $\true$, $T_{cond}$ puts a token in a place called $GuardOk$ that, in turn, allows to 
initiate the action execution process that is sequentially realized for all $R$-facts from $F^{-}$ or $F^{+}$ in \actname{Act}.

\begin{figure}[t!]
\subfigure[Expansion of the $i$-th deletion component in the net of Figure~\ref{fig:update-net}\label{fig:update-net-expansion-del}]{
\scalebox{.8}{
\begin{tikzpicture}
 [x=1cm,y=1cm,node distance=1.4cm,very thick]
 
 \node[
  place
 ] (pin) {};

 \node[
    place,
    right=of pin,
    ] (pdone) {};
\node[below right=-1mm of pdone,yshift=2mm] {$\mathit{DoneD_i}$};

 \node[
  transition,
  above=of pdone,
  yshift=-2mm,
  label=left:$\mathit{ExistsD_i}$
  ] (te) {};
\node[right=0mm of te] {\phigh};

 \node[
  transition,
  below=of pdone,
  yshift=3mm,
  label=left:$\mathit{NotExistsD_i}$
  ] (tn) {};
\node[right=0mm of tn] {\plow};
 
 \node[
  place,
  right=of pdone
  ] (pout) {};

\node[
  place,
  above=of te,
    yshift=-5mm,
  label={[yshift=-1mm]above:$R_{d,i}$}
  ] (pr) {};

  \node[right=3mm of pdone] {$\circ$};
  \node[left=3mm of pdone] {$\circ$};
  \node[right=3mm of pr] {$\circ$};
  \node[left=3mm of pr] {$\circ$};

\begin{pgfonlayer}{background}

 \node[
  noopelem,
  fit=(pdone),
    minimum width=35mm,
  inner sep=2mm,
  label={[xshift=-5mm,yshift=-1mm]above right:\begin{tabular}{@{}c@{}}no-op\\places\end{tabular}},
  ] (noop) {};

 \node[
  relationselem,
  fit=(pr),
  minimum width=35mm,
  minimum height=10mm,
  inner sep=1.5mm,
  label=left:relation places
  ] (noop) {};
\end{pgfonlayer}

\path[-stealth', thick]
  (pin)
  edge[out=90,in=180] node[left] {$\tup{\vec{z},\vec{x}}$}
  ($(te.south west)+(0,1mm)$)
  ($(te.south east)+(0,1mm)$)
  edge[out=0,in=90] node[right] {$\tup{\vec{z},\vec{x}}$}
  (pout)
  (pr)
  edge node[below left] {$\vec{y}_{d,i}$}
  (te)
  (te)
  edge node[above left] {$\tup{\true}$}
  (pdone)
  (pin)
  edge[out=-90,in=180] node[left] {$\tup{\vec{z},\vec{x}}$}
  ($(tn.north west)-(0,1mm)$)
  ($(tn.north east)-(0,1mm)$)
  edge[out=0,in=-90] node[right] {$\tup{\vec{z},\vec{x}}$}
  (pout)
  (tn)
  edge node[below left] {$\tup{\false}$}
  (pdone)
  ;
 
\end{tikzpicture}
}
}
\hfill
\subfigure[Expansion of the $i$-th addition component in the net of Figure~\ref{fig:update-net}\label{fig:update-net-expansion-add}]{
\scalebox{.8}{
\begin{tikzpicture}
 [x=1cm,y=.5cm,node distance=1.4cm,very thick]
 
 \node[
  place
 ] (pin) {};

 \node[
    place,
    right=of pin,
    ] (pdone) {};
\node[below right=-1mm of pdone,yshift=2mm] {$\mathit{DoneA_i}$};

 \node[
  transition,
  above=of pdone,
  yshift=-2mm,
  label=left:$\mathit{ExistsA_i}$
  ] (te) {};
\node[right=0mm of te] {\phigh};

 \node[
  transition,
  below=of pdone,
  yshift=3mm,
  label=left:$\mathit{NotExistsA_i}$
  ] (tn) {};
\node[right=0mm of tn] {\plow};
 
 \node[
  place,
  right=of pdone
  ] (pout) {};

\node[
  place,
  above=of te,
  yshift=-5mm,
  label={[yshift=-1mm]above:$R_{a,i}$}
  ] (pr) {};

  \node[right=3mm of pdone] {$\circ$};
  \node[left=3mm of pdone] {$\circ$};
  \node[right=3mm of pr,yshift=2mm] {$\circ$};
  \node[left=3mm of pr,yshift=2mm] {$\circ$};

\begin{pgfonlayer}{background}

 \node[
  noopelem,
  fit=(pdone),
  minimum width=35mm,
  inner sep=2mm,
  label={[xshift=-5mm,yshift=-1mm]above right:\begin{tabular}{@{}c@{}}no-op\\places\end{tabular}},
  ] (noop) {};

 \node[
  relationselem,
  fit=(pr),
  minimum width=35mm,
  minimum height=10mm,
  inner sep=1.5mm,
  label=left:relation places
  ] (noop) {};
\end{pgfonlayer}

\path[-stealth', thick]
  (pin)
  edge[out=90,in=180] node[left] {$\tup{\vec{z},\vec{x}}$}
  ($(te.south west)+(0,1mm)$)
  ($(te.south east)+(0,1mm)$)
  edge[out=0,in=90] node[right] {$\tup{\vec{z},\vec{x}}$}
  (pout)
  (te)
  edge node[above left] {$\tup{\false}$}
  (pdone)
  (pin)
  edge[out=-90,in=180] node[left] {$\tup{\vec{z},\vec{x}}$}
  ($(tn.north west)-(0,1mm)$)
  ($(tn.north east)-(0,1mm)$)
  edge[out=0,in=-90] node[right] {$\tup{\vec{z},\vec{x}}$}
  (pout)
  (tn)
  edge node[below left] {$\tup{\true}$}
  (pdone)
  ;
  
\path[-,thick]
  (pr)
  edge node[below left] {$\vec{y}_{a,i}$}
  (te)
;

\draw[-stealth',thick,rounded corners=35pt]
  ($(tn.south east)+(0,1mm)$)
  -| node[above] {$\vec{y}_{a,i}$}
  ($(pout.east)+(4mm,1cm)$)
  |-
  (pr)
;
 
\end{tikzpicture}
}

}

\caption{Expansion of deletion and addition nets from Figure~\ref{fig:update-net}}  
\label{fig:update-net-expansion}
\end{figure}
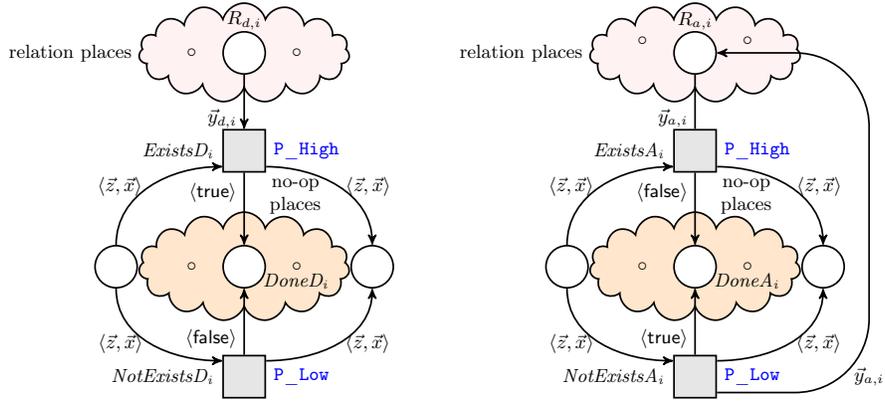

We first proceed with deleting all the $R_{d,i}$-facts from $F^{-}$ (i.e., facts of the form $R_{d,i}(\vec{y}_{d,i})$). This process is sequntialized and at each of its step the net models the deletion of only one $R_{d,i}$-fact. 
Specifically, the deletion of each $R_{d,i}$-fact (see Figure~\ref{fig:update-net-expansion-del}) is
realized by 
a pair of prioritized transitions \transition{ExistsD$_{i}$} and \transition{NotExistsD$_{i}$} and one auxiliary place $DoneD_i$, and 
is analogous to the example in Figure~\ref{fig:dbnets-add-trick-2}. 
After all the $R$-facts from $F^{-}$ have been deleted, 
the net switches to performing the insertion of $R$-facts from $F^{+}$. We omit 
the details of the addition process as it can be defined analogously to the one from the bonus change example and refer to Figure~\ref{fig:update-net-expansion-add}. As soon as all $R$-facts are added, the update net completes its work by putting a token into a place called $Updated$.

\subsection{Checking Integrity Constraints and Generating a New Marking}
\label{sec:encoding-constraints}

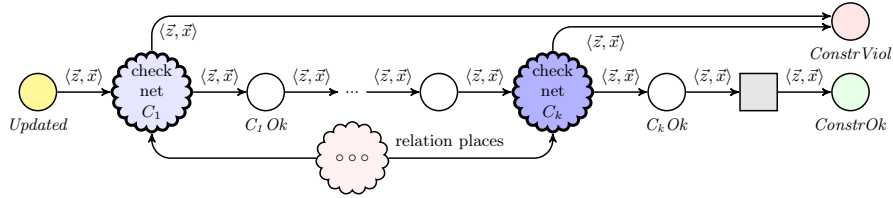
\begin{figure}[t!]
\resizebox{0.99\textwidth}{!}{
\begin{tikzpicture}
 [x=1cm,y=1cm,node distance=1cm,very thick,font=\footnotesize]
 
\node[
  updatedplace,
  label=below:$\mathit{Updated}$
  ] (pupdated) {};
    
\node[
  netelem,
  fill=blue!10,
  right=of pupdated,
    minimum width=1.5cm,
  minimum height=1.5cm,
  label=center:\begin{tabular}{@{}c@{}}check\\net\\$C_1$\end{tabular}
  ] (check1) {};

\node[
  place,
  right=of check1,
  label=below:$\mathit{C_1Ok}$
  ] (pc1out) {};
  
\node[right=of pc1out] (dots) {...};

\node[
  place,
  right=of dots
  ] (pckin) {};
  
\node[
  netelem,
  fill=blue!30,
  right=of pckin,
  minimum width=1.5cm,
    minimum height=1.5cm,
  label=center:\begin{tabular}{@{}c@{}}check\\net\\$C_k$\end{tabular}
  ] (checkk) {};

\node[
  place,
  right=of checkk,
  label=below:$\mathit{C_kOk}$
  ] (pckout) {};
 
 \node[
  transition,
  right=of pckout
  ] (tok) {};

\node[
  constrokplace,
  right=of tok,
  label=below:$\mathit{ConstrOk}$
  ] (pok) {};

\node[
  violplace,
  above=of pok,
  yshift=-4mm,
  label=below:$\mathit{ConstrViol}$
  ] (pviol) {};

\node[
  relationselem,
  below=of dots,
  yshift=6mm,
  label={[yshift=3mm]right:relation places}
  ] (relations) {$\circ \circ \circ$};

\path[arc]
(pupdated)
  edge node[above] {$\tup{\vec{z},\vec{x}}$}
(check1)
(check1)
  edge node[above] {$\tup{\vec{z},\vec{x}}$}
(pc1out)
(pc1out)
  edge node[above] {$\tup{\vec{z},\vec{x}}$}
(dots)
(dots)
  edge node[above] {$\tup{\vec{z},\vec{x}}$}
(pckin)
(pckin)
  edge node[above] {$\tup{\vec{z},\vec{x}}$}
(checkk)
(checkk)
  edge node[above] {$\tup{\vec{z},\vec{x}}$}
(pckout)
(pckout)
  edge node[above] {$\tup{\vec{z},\vec{x}}$}
(tok)
(tok)
  edge node[above] {$\tup{\vec{z},\vec{x}}$}
(pok);

\draw[arc,rounded corners=10pt]
(relations)
    -| 
(check1);

\draw[arc,rounded corners=10pt]
(relations)
    -| 
(checkk);

\draw[arc,rounded corners=10pt]
(check1)
    |- node[below,xshift=5mm,yshift=-.2mm] {$\tup{\vec{z},\vec{x}}$}
($(pviol.west)+(0,1mm)$);

\draw[arc,rounded corners=10pt]
(checkk)
    |- node[right,xshift=5mm,yshift=-3mm] {$\tup{\vec{z},\vec{x}}$}
($(pviol.west)-(0,1mm)$);

\end{tikzpicture}
}
\caption{Expansion of the check constraint net from Figure~\ref{fig:cpn-encoding}}  
\label{fig:check-constraint-net}
\end{figure}

Let us now remind that the relational schema of $\dbn_\tau$ is equipped with three types of integrity constraints: primary keys, foreign keys and domain  constraints. When the first and the last one could be relatively easy to check during the update phase, assuming that the computation results are 
accumulated in arc inscriptions analogously to the binding net in Figure~\ref{fig:phase1}\footnote{Both primary keys 
and domain constraints can be violated when a tuple is about to be inserted into a table. Specifically, to guarantee that primary keys are respected, it is enough to check with $ExistsA_i$ whether there is a token in $R_{a,i}$ that has the same primary key value, and, if so, cancel the computation process. 
In the case of check constraints, one may insert a third transition that has a normal priority and that will be fired whenever one of the values we want to insert is not falling into the allowed range. Firing of this transition will have the same consequences as in the case of primary keys.}, the process of managing updates in the presence of arbitrary many foreign key dependencies is quite involved. 
To manage it correctly we first perform the updates and only then check whether the generated marking 
represents a database instance that satisfies all the integrity constraints contained in the persistence layer of $\dbn_\tau$. A \nucpn representing the check constraint phase is depicted in Figure~\ref{fig:check-constraint-net}.
The net works as follows: it consequently runs small nets for verifying the integrity of constraints and, in case of violation, puts a token in a special place called $ConstrViol$. As soon as there is at least one token in 
$ConstrViol$ place, the big net in Figure~\ref{fig:cpn-encoding} terminates the constraint checking process and switches to the phase 4.(b) (that is, runs the undo net) explained in the beginning of this section. 
For ease of presentation we assume that every relation $R/n_r$ from $\schema$ of $\dbn_\tau$ 
will have the following look: the first $k$ attributes form a primary key, while the rest of $n_r-k+1$ attributes
can be unconstrained or bounded by domain constraints. Moreover, if $R$ is referencing some other relation $S$, then among these  $n_r-k+1$ attributes we reserve the last $m$ such that $\FKindex{S}{\set{1,\ldots,m}}{R}{\set{n_r-m,\ldots,n_r}}$.

\begin{figure}[t!]
\centering
\subfigure[Expansion for a primary key constraint $C_i$ indicating that the first $k$ components of relation $R$ with arity $n_r \geq k$ form a key for $R$. In the figure, $\vec{y}$ and $\vec{w}$ are tuples containing $n_r$ variables.\label{fig:check-constraint-net-pk}] 
{
\resizebox{.47\textwidth}{!}{
\begin{tikzpicture}
 [x=1cm,y=1cm,node distance=1.4cm,very thick,font=\footnotesize]

 \node[
  place
 ] (pin) {};
 
 \node[
    place,
    right=of pin,
    label={[yshift=1mm]below:$R$}
    ] (pr) {};

 \node[
  transition,
  above=of pr,
  yshift=-5mm,
  label=left:$\mathit{RepeatedKey}$
  ] (tr) {};
\node[below right=-1mm of tr,yshift=2mm] {\phigh};

\node[above=0mm of tr] {  
  $\mathbf{[}{\color{mgreen}
    \bigwedge_{j =\overline{1,k}}
    y_j = w_j
    \land
    \bigvee_{j = \overline{k+1,n_r}}
    y_j \neq w_j
  }\mathbf{]}$
};

 \node[
  transition,
  below=of pr,
  yshift=6mm,
  label=left:$\mathit{NoRepeatedKey}$
  ] (tn) {};
\node[above right=0mm of tn,xshift=-2mm,yshift=-1mm] {\plow};
 
 \node[
  place,
  right=of tn,
  label=above:$\mathit{C_iOk}$
  ] (pout) {};

 \node[
  violplace,
  right=of tr,
  label=below:$\mathit{ConstrViol}$
  ] (pviol) {};

  \node[right=3mm of pr] {$\circ$};
  \node[left=3mm of pr] {$\circ$};

\begin{pgfonlayer}{background}
 \node[
  relationselem,
  fit=(pr),
  minimum width=25mm,
  minimum height=15mm,
  inner sep=1.2mm,
  label={[yshift=-.6mm]right:relation places},
  ] (relations) {};
\end{pgfonlayer}

\path[arc]
  (pin)
  edge[out=70,in=180] node[left,xshift=-1mm] {$\tup{\vec{z},\vec{x}}$}
  ($(tr.south west)+(0,1mm)$)
  (tr)
  edge node[above] {$\tup{\vec{z},\vec{x}}$}
  (pviol)
  (pin)
  edge[out=-70,in=180] node[left,xshift=-1mm] {$\tup{\vec{z},\vec{x}}$}
  ($(tn.north west)-(0,1mm)$)
  (tn)
  edge node[below] {$\tup{\vec{z},\vec{x}}$}
  (pout)
  ;
  
\path[readarc]
  ($(pr.north)-(1mm,0)$)
  edge node[left,pos=0.75] {$\vec{y}$}
  ($(tr.south)-(1mm,0)$)
  ($(pr.north)+(1mm,0)$)
  edge node[right,pos=0.75] {$\vec{w}$}
  ($(tr.south)+(1mm,0)$)
;
\end{tikzpicture}
}
}
\subfigure[Expansion for a foreign key constraint $C_i$ indicating that the last $m$ components of relation $R$ with arity $n_r \geq m$ reference the last $m$ components of relation $S$ with arity $n_s \geq m$. In the figure, $\vec{y}$ and $\vec{w}$ are tuples containing $n_r$ and $n_s$ variables.\label{fig:check-constraint-net-fk}] {
\resizebox{.47\textwidth}{!}{
\begin{tikzpicture}
 [x=.5cm,y=1cm,node distance=1.4cm,very thick,font=\footnotesize]
 
 \node[
  place
 ] (pin) {};

 \node[right=of pin] (somerel) {$\circ$};

 \node[
    place,
    left=1mm of somerel,
    label=left:$R$
    ] (pr) {};
 \node[
    place,
    right=1mm of somerel,
    label=right:$S$
    ] (ps) {};
 
 \node[
  transition,
  above=of somerel,
    yshift=-4mm,
  label=left:$\mathit{FKExists}$
  ] (tr) {};
\node[below right=-1mm of tr,yshift=2mm] {\phigh};

\node[above=0mm of tr] {  
  $\mathbf{[}{\color{mgreen}
    \bigwedge_{j = \overline{1,m}}
    y_{n_r-m+j} = w_j
  }\mathbf{]}$
};

 \node[
  transition,
  below=of somerel,
  yshift=5mm,
  label=left:$\mathit{FKNotExists}$
  ] (tn) {};
\node[above right=0mm of tn,xshift=-2mm,yshift=-1mm] {\plow};
 
 \node[
  violplace,
  right=of tn,
  label=above:$\mathit{ConstrViol}$
  ] (pout) {};

 \node[
  place,
  right=of tr,
  label=below:$\mathit{C_iOk}$
  ] (pviol) {};

\begin{pgfonlayer}{background}
 \node[
  relationselem,
  fit=(somerel),
  minimum width=31mm,
  inner sep=2mm,
   label={[yshift=-.6mm]right:relation places},
  ] (relations) {};
\end{pgfonlayer}

\path[arc]
  (pin)
  edge[out=70,in=180] node[left,xshift=-1mm] {$\tup{\vec{z},\vec{x}}$}
  ($(tr.south west)+(0,1mm)$)
  (tr)
  edge node[above] {$\tup{\vec{z},\vec{x}}$}
  (pviol)
   (pin)
  edge[out=-70,in=180] node[left,xshift=-1mm] {$\tup{\vec{z},\vec{x}}$}
  ($(tn.north west)-(0,1mm)$)
  (tn)
  edge node[below] {$\tup{\vec{z},\vec{x}}$}
  (pout)
  ;
 
\path[readarc]
 (pr)
  edge node[left,pos=0.6] {$\vec{y}$}
  (tr)
  (ps)
  edge node[right,pos=0.6] {$\vec{w}$}
  (tr)
;
\end{tikzpicture}
}
}
\caption{Expansion of the check net from Figure~\ref{fig:check-constraint-net} in the case of  key constraints}
\end{figure}
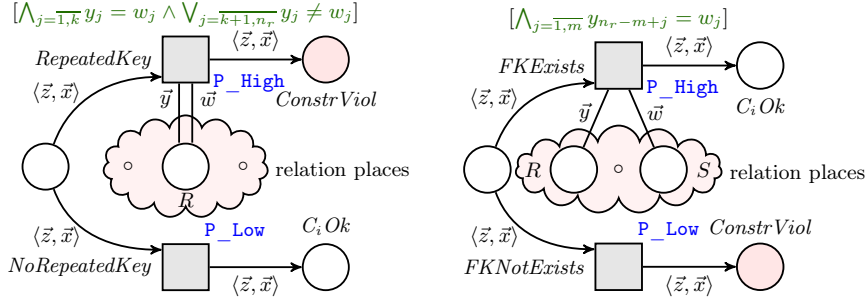

The constraint checking process starts with verifying that all the updates performed using the net in
Figure~\ref{fig:update-net} are satisfying primary key constraints $C_i$. 
This is done by sequentially running small check nets in Figure~\ref{fig:check-constraint-net-pk}, where
the constraint integrity is verified for some relation $R$ by a pair of prioritized transitions $RepeatedKey$ (high priority) and $NoRepeatedKey$ (low priority). Note that $RepeatedKey$ accesses the content of $R$
with two read arcs and using the guard assigned to it verifies whether there exist two tokens, such that their first $k$ values coincide and in the rest of $n_r-k+1$ values there is at least one distinct pair of values. 
The satisfaction of such guard would mean that, essentially, we have inserted a token in $R$ whose primary 
key values were not unique. Firing of $RepeatedKey$ will produce one token in $ConstViol$ and terminate the run of the check constraint net.

The next type of constraints to verify is the foreign key dependency. Analogously to the previous case, 
we successively run small check nets like the one in Figure~\ref{fig:check-constraint-net-fk} and 
in each of them control that $R$ correctly references $S$ (that is, there are no tuples in $R$ that do not depend on any tuple in $S$).  
This is realized with two prioritized transitions $FKExists$ (high priority) and $FKNotExists$ (low priority). 
The first one, as the name suggests, checks whether the dependency between $R$ and $S$ is preserved for all the tokens in the corresponding relation places. 
$FKExists$ makes use of the guard attached to it that performs pairwise comparison of $m$ last values of a token from $R$ to $m$ first values of a token from $S$. 
If the guard is not satisfied, then the dependency relation between $R$ and $S$
has been violated and one fires $FKNotExists$ so as to terminate the constraint checking process.

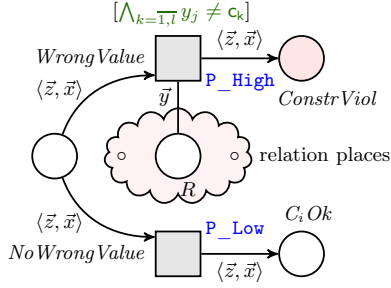
\begin{wrapfigure}[16]{l}{.5\textwidth}
\vspace*{-.2cm}
\resizebox{.45\textwidth}{!}{
\begin{tikzpicture}
 [x=1cm,y=1cm,node distance=1.2cm,very thick,font=\footnotesize]
 
 \node[
  place
 ] (pin) {};

 \node[
    place,
    right=of pin,
    label={[yshift=1mm,xshift=1.5mm]below:$R$}
    ] (pr) {};
 
 \node[
  transition,
  above=of pr,
  yshift=-4mm,
  label=left:$\mathit{WrongValue}$
  ] (tr) {};
\node[below right=-1mm of tr,yshift=2mm] {\phigh};

\node[above=0mm of tr] {{  
  $\mathbf{[}{\color{mgreen}
    \bigwedge_{k = \overline{1,l}}
    y_j \neq \cval{c_k}
  }\mathbf{]}$}
};

 \node[
  transition,
  below=of pr,
  yshift=4mm,
  label=left:$\mathit{NoWrongValue}$
  ] (tn) {};
\node[above right=-1mm of tn,yshift=-2mm] {\plow};
 
 \node[
  place,
  right=of tn,
  label={[xshift=1mm]above:$\mathit{C_iOk}$}
  ] (pout) {};

 \node[
  violplace,
  right=of tr,
  label={[xshift=4mm]below:$\mathit{ConstrViol}$}
  ] (pviol) {};

  \node[right=3mm of pr] {$\circ$};
  \node[left=3mm of pr] {$\circ$};

\begin{pgfonlayer}{background}
 \node[
  relationselem,
  fit=(pr),
  minimum width=23mm,
  inner sep=1mm,
  label={[yshift=0mm]right:relation places},
  ] (relations) {};
\end{pgfonlayer}

\path[arc]
  (pin)
  edge[out=70,in=180] node[left,xshift=-1mm] {$\tup{\vec{z},\vec{x}}$}
  ($(tr.south west)+(0,1mm)$)
  (tr)
  edge node[above] {$\tup{\vec{z},\vec{x}}$}
  (pviol)
  (pin)
  edge[out=-70,in=180] node[left,xshift=-1mm] {$\tup{\vec{z},\vec{x}}$}
  ($(tn.north west)-(0,1mm)$)
  (tn)
  edge node[below] {$\tup{\vec{z},\vec{x}}$}
  (pout)
  ;

\path[readarc]
    (pr)
  edge node[left,pos=0.7] {$\vec{y}$}
  (tr); 
\end{tikzpicture}
}
\caption{Expansion of the check net from Figure~\ref{fig:check-constraint-net} in the case of a domain constraint $C_i$ indicating that the $j$-th component of relation $R$ with arity $n_r \geq j$ must contain a value that belongs to $\set{\cval{c_1},\ldots,\cval{c_l}}$; In the figure, $\vec{y}$ is a  tuple containing $n_r$ variables.}
\label{fig:check-constraint-net-dom}
\end{wrapfigure}
The last series of constraints to be checked is the one of domain constraints. The net in Figure~\ref{fig:check-constraint-net-dom} employs two prioritized transitions, $WrongValue$ and 
$NoWrongValue$, to verify whether all the tuples inserted into $R$ had correct values. 
First, using the guard of $WrongValue$, we check whether there is at least one value that breaks the integrity of the domain constraint $C_i$ of $R$. If $WrongValue$ fires, the process is terminated by putting a token into $ConstrViol$. Otherwise, $NoWrongValue$ is executed and the constraint checking process continues.

Now let us show how the computation of all the effects of \transition{T} is finished and a new marking is generated. 
If one of the constraints has been violated, we have to roll back all the effects pushed using
the net in Figure~\ref{fig:update-net}. 
To do so, we employ the net in Figure~\ref{fig:undo-net} that reverts the update process by
first canceling all the additions, and then canceling all the deletions. Let us briefly explain 
how the rollback process is performed for each component net.

\begin{figure}[h!]
\centering
\vspace*{-2mm}
\resizebox{\textwidth}{!}{
\begin{tikzpicture}
  [x=1cm,y=1cm,node distance=1cm,thick]
  
\node[
  violplace,
  label=below:$\mathit{ConstrViol}$
  ] (pviol) {};

\node[
  netelem,
  label=center:
      \begin{tabular}{@{}c@{}}revert\\add net\\($R_{a,1}(\vec{y}_{a,1})$)\end{tabular},
  right=of pviol,
  fill=mgreen!10,
  minimum width=2.2cm,
  minimum height=1.8cm,
  ] (revadd1) {};
  
\node[
  place,
  right=of revadd1
  ] (prevadd1) {};

\node[
  right=of prevadd1
  ] (dotrevadd) {...};
  
 \node[
  place,
  right=of dotrevadd
  ] (prevaddn) {}; 
  
\node[
  netelem,
  label=center:
      \begin{tabular}{@{}c@{}}revert\\del net\\($R_{d,1}(\vec{y}_{d,1})$)\end{tabular},
  right=of prevaddn,
  fill=red!10,
  minimum width=2.2cm,
  minimum height=1.8cm,
  ] (revdel1) {};
  
\node[
  place,
  right=of revdel1
  ] (prevdel1) {};

\node[
  right=of prevdel1
  ] (dotrevdel) {...};
  
\node[
  dorollbackplace,
  label=below:$\mathit{DoRollback}$,
  right=of dotrevdel
  ] (prb) {};

\node[
  relationselem,
  label={[yshift=-3mm]left:relation places},
  yshift=-6mm,
  above=of prevaddn
  ] (relations) {$\circ \circ \circ$};
  
\node[
  noopelem,
  label={[yshift=3mm]left:no-op places},
  yshift=6mm,
  below=of prevaddn
  ] (noop) {$\circ \circ \circ$};

\path[-stealth']
(pviol)
  edge node[above] {$\tup{\vec{z},\vec{x}}$}   
(revadd1)
(revadd1)
  edge node[above] {$\tup{\vec{z},\vec{x}}$}   
(prevadd1)
(prevadd1)
  edge node[above] {$\tup{\vec{z},\vec{x}}$}   
(dotrevadd)
(dotrevadd)
  edge node[above] {$\tup{\vec{z},\vec{x}}$}   
(prevaddn)
(prevaddn)
  edge node[above] {$\tup{\vec{z},\vec{x}}$}   
(revdel1)
(revdel1)
  edge node[above] {$\tup{\vec{z},\vec{x}}$}   
(prevdel1)
(prevdel1)
  edge node[above] {$\tup{\vec{z},\vec{x}}$}   
(dotrevdel)
(dotrevdel)
  edge node[above] {$\tup{\vec{z},\vec{x}}$}   
(prb)
;

\draw[-stealth',rounded corners=10pt]
(relations.west)
  -| node[left] {$\vec{y}_{a,1}$}
(revadd1.north);
\draw[-stealth',rounded corners=10pt]
(noop.west)
  -| node[xshift=1mm,left] {$\tup{b}$}
(revadd1.south);
\draw[-stealth',rounded corners=10pt]
(revdel1.north)
  |- node[right] {$\vec{y}_{d,1}$}
(relations.east);
\draw[-stealth',rounded corners=10pt]
(noop.east)
  -| node[xshift=-1mm,right] {$\tup{b}$}
(revdel1.south)
;


\end{tikzpicture}  
}
\caption{Expansion of the undo net from Figure~\ref{fig:cpn-encoding}}
\label{fig:undo-net}
\end{figure}
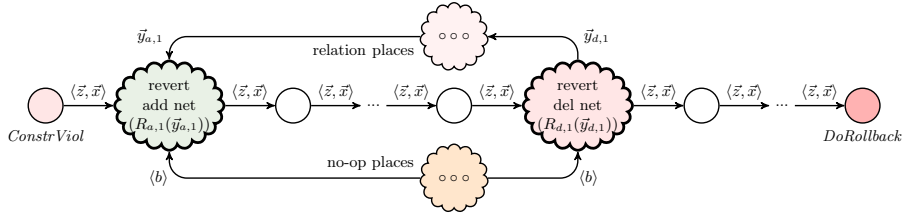

\begin{figure}[t!]
\subfigure[Expansion of the $i$-th revert addition component. \label{fig:revert-add-net}]{
\scalebox{.8}{
\begin{tikzpicture}
 [x=1cm,y=1cm,node distance=1.4cm,very thick]
 
 \node[
  place
 ] (pin) {};

 \node[
    place,
    right=of pin,
    ] (pdone) {};
\node[below right=-2mm of pdone,yshift=1mm] {$\mathit{DoneA_i}$};

 \node[
  transition,
  above=of pdone,
        yshift=-1mm,
  label=left:$\mathit{DoRevertA_i}$
  ] (te) {};

 \node[
  transition,
  below=of pdone,
      yshift=2mm,
  label=left:$\mathit{SkipRevertA_i}$
  ] (tn) {};
 
 \node[
  place,
  right=of pdone
  ] (pout) {};

\node[
  place,
  above=of te,
  yshift=-4mm,
  label={[yshift=-1mm]above:$R_{a,i}$}
  ] (pr) {};

  \node[right=3mm of pdone,yshift=2mm] {$\circ$};
  \node[left=3mm of pdone,yshift=2mm] {$\circ$};
  \node[right=3mm of pr,yshift=2mm] {$\circ$};
  \node[left=3mm of pr,yshift=2mm] {$\circ$};

\begin{pgfonlayer}{background}

 \node[
  noopelem,
  fit=(pdone),
  minimum width=33mm,
  inner sep=2mm,
  label={[xshift=-7mm]above right:\begin{tabular}{@{}c@{}}no-op\\places\end{tabular}},
  ] (noop) {};

 \node[
  relationselem,
  fit=(pr),
minimum width=30mm,
  inner sep=2mm,
  label=below left:relation places
  ] (noop) {};
\end{pgfonlayer}

\path[-stealth', thick]
  (pin)
  edge[out=90,in=180] node[left] {$\tup{\vec{z},\vec{x}}$}
  ($(te.south west)+(0,1mm)$)
  ($(te.south east)+(0,1mm)$)
  edge[out=0,in=90] node[right] {$\tup{\vec{z},\vec{x}}$}
  (pout)
  (pr)
  edge node[below left] {$\vec{y}_{a,i}$}
  (te)
  (pdone)
  edge node[above left] {$\tup{\true}$}
  (te)
  (pin)
  edge[out=-90,in=180] node[left] {$\tup{\vec{z},\vec{x}}$}
  ($(tn.north west)-(0,1mm)$)
  ($(tn.north east)-(0,1mm)$)
  edge[out=0,in=-90] node[right] {$\tup{\vec{z},\vec{x}}$}
  (pout)
  (pdone)
  edge node[below left] {$\tup{\false}$}
  (tn)
  ;
 
\end{tikzpicture}
}

}
\hfill
\subfigure[Expansion of the $i$-th deletion component.\label{fig:revert-del-net}]{
\scalebox{.8}{
\begin{tikzpicture}
 [x=1cm,y=1cm,node distance=1.4cm,very thick]

  \node[
  place
 ] (pin) {};

 \node[
    place,
    right=of pin,
    ] (pdone) {};
\node[below right=-2mm of pdone,yshift=1mm] {$\mathit{DoneD_i}$};

 \node[
  transition,
  above=of pdone,
      yshift=-1mm,
  label=left:$\mathit{DoRevertD_i}$
  ] (te) {};

 \node[
  transition,
  below=of pdone,
    yshift=2mm,
  label=left:$\mathit{SkipRevertD_i}$
  ] (tn) {};
 
 \node[
  place,
  right=of pdone
  ] (pout) {};

\node[
  place,
  above=of te,
    yshift=-4mm,
  label={[yshift=-1mm]above:$R_{d,i}$}
  ] (pr) {};

  \node[right=3mm of pdone,yshift=2mm] {$\circ$};
  \node[left=3mm of pdone,yshift=2mm] {$\circ$};
  \node[right=3mm of pr,yshift=2mm] {$\circ$};
  \node[left=3mm of pr,yshift=2mm] {$\circ$};

\begin{pgfonlayer}{background}

 \node[
  noopelem,
  fit=(pdone),
  minimum width=33mm,
  inner sep=2mm,
  label={[xshift=-7mm]above right:\begin{tabular}{@{}c@{}}no-op\\places\end{tabular}},
  ] (noop) {};

 \node[
  relationselem,
  fit=(pr),
  minimum width=30mm,
  inner sep=2mm,
  label=below left:relation places
  ] (noop) {};
\end{pgfonlayer}

\path[arc]
  (pin)
  edge[out=90,in=180] node[left] {$\tup{\vec{z},\vec{x}}$}
  ($(te.south west)+(0,1mm)$)
  ($(te.south east)+(0,1mm)$)
  edge[out=0,in=90] node[right] {$\tup{\vec{z},\vec{x}}$}
  (pout)
  (te)
  edge node[below left] {$\vec{y}_{d,i}$}
  (pr)
  (pdone)
  edge node[above left] {$\tup{\true}$}
  (te)
  (pin)
  edge[out=-90,in=180] node[left] {$\tup{\vec{z},\vec{x}}$}
  ($(tn.north west)-(0,1mm)$)
  ($(tn.north east)-(0,1mm)$)
  edge[out=0,in=-90] node[right] {$\tup{\vec{z},\vec{x}}$}
  (pout)
  (pdone)
  edge node[below left] {$\tup{\false}$}
  (tn)
  ;
  
\end{tikzpicture}
}

}

\caption{Expansion of revert deletion and addition nets from Figure~\ref{fig:undo-net}}  
\label{fig:revert-nets}
\end{figure}
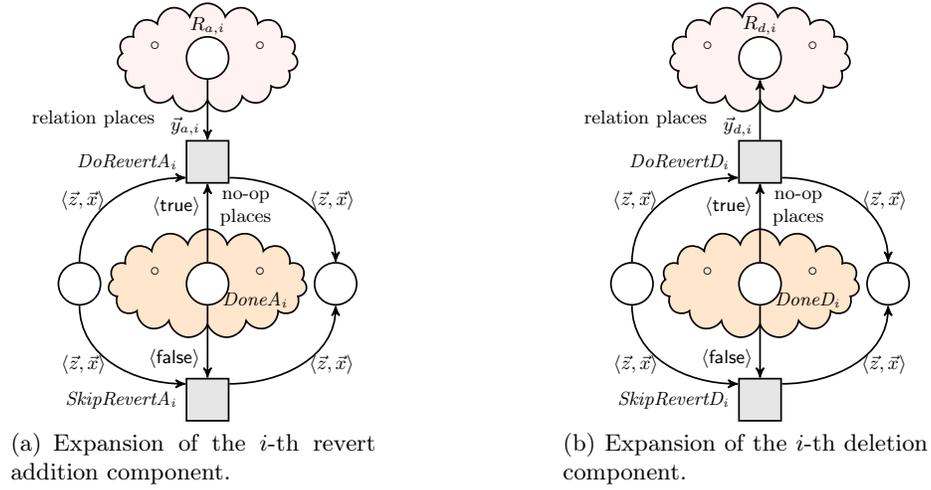

We start by removing all the tuples that have been successfully added to relation places following 
the definition of $\actname{Act}$. 
The net in Figure~\ref{fig:revert-add-net} shows how to revert
the result of inserting $R_{a,i}$-fact from $F^{+}$ (i.e., a fact of the form $R_{a,i}(\vec{y}_{a,i})$).
If a fact has been added, that is, there is a token with value $\tup{\true}$ in $DoneA_i$, then 
the net removes it by firing $DoRevertA_i$. 
Otherwise, if the fact has not been added, that is, there is a token with value $\tup{\false}$ in $DoneA_i$, then the net proceeds without reverting by firing $SkipRevertA_i$. 
Then, for each $R_{d,i}$-fact, we go on with adding all the tuples that have been deleted by using the net depicted in Figure~\ref{fig:revert-del-net}. 
The update reverting processis analogous to the one dealing with reverted additions, but with only one exception: whenever 
$DoneD_i$ has a $\tup{\true}$ token, then we put the deleted tuple (specified in $F^{-}$)  back into $R_{d,i}$. 
Note that every revert deletion or addition net removes a token from a corresponding auxiliary no-op place.

As soon as all the operations of $\actname{Act}$ have been undone and all the corresponding tokens have been withdrawn from relation places, the net places a token in $DoRollBack$ (cf. Figure~\ref{fig:cpn-encoding}) and allows us to fire a transition called $T_{rollback}$ that implements the generation of the tokens in the postset corresponding to the rollback flow of \transition{T}.

\begin{figure}[t!]
\centering
\resizebox{.8\textwidth}{!}{
\begin{tikzpicture}
 [x=1cm,y=.5cm,node distance=2.5cm,very thick]
 
\node[
  constrokplace,
  yshift=26mm,
  label=below:$\mathit{ConstrOk}$
  ] (pok) {};
  
\node[
  transition,
  label=below:$\mathit{EmptyNoOpPlaces}$,
  right=of pok
  ] (t) {};

\node[
  docommitplace,
  label=below:$\mathit{DoCommit}$,
  right=of t
  ] (pc) {};

\node[above=1.5cm of t] (ref) {};

\node[
  place,
  left=4mm of ref,
  yshift=-.7mm,
  label={[yshift=-1.2mm]above:$DoneD_{n_d}$}
  ] (pdn) {};
  
\node[left=3mm of pdn] (ddots) {...};

\node[
  place,
  left=3mm of ddots,
  yshift=-1.3mm,
  label={[yshift=-1mm]above:$DoneD_1$}
  ] (pd1) {};

\node[
  place,
  right=3mm of ref,
  yshift=-2mm,
  label={[yshift=-1mm]above:$DoneA_1$}
  ] (pa1) {};
  
\node[right=3mm of pa1] (adots) {...};

\node[
  place,
  right=4mm of adots,
  yshift=-.3mm,
  label={[yshift=-1.2mm]above:$DoneA_{n_a}$}
  ] (pan) {};

\begin{pgfonlayer}{background}

 \node[
  noopelem,
  fit=(ref),
  minimum width=102mm,
  inner sep=5mm,
  cloud puffs = 60,
  label=right:no-op places,
  ] (noop) {};
\end{pgfonlayer}

\path[arc]
(pok)
  edge node[below] {$\tup{\vec{z},\vec{x}}$}
(t)
(pd1)
  edge[bend right=9] node[xshift=-1mm, left,pos=0.7] {$\tup{b_{d,1}}$}
(t)
(pdn)
  edge node[left,pos=0.6,xshift=1mm] {$\tup{b_{d,n_d}}$}
(t)
(pa1)
  edge node[right,pos=0.6,xshift=-1mm] {$\tup{b_{a,1}}$}
(t)
(pan)
  edge[bend left=9] node[xshift=2mm, right,pos=0.7] {$\tup{b_{a,n_a}}$}
(t)
(t)
  edge node[below] {$\tup{\vec{z},\vec{x}}$}
(pc)
;

\end{tikzpicture}
}
\caption{Expansion of the consume net from Figure~\ref{fig:cpn-encoding}\label{fig:consume-net}}
\end{figure}
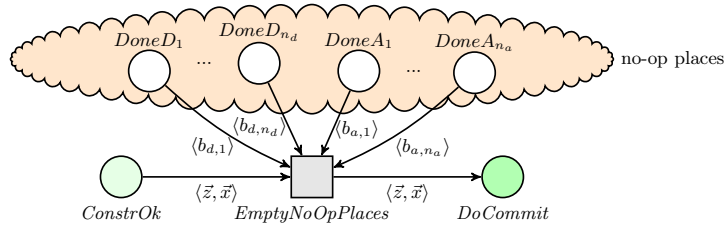

If the check constraint net's work has not been interrupted an the token was placed in $ConstrOk$ (cf. Figure~\ref{fig:cpn-encoding}), then we proceed with the consume net (cf. Figure~\ref{fig:consume-net})
that removes all tokens from the auxiliary no-op places and places a token into $DoCommit$. 
This, in turn, allows $\cpn_\tau$ to execute $T_{commit}$ that populates tokens in the postset 
corresponding to the normal flow of \transition{T}.

\makeatletter
\newcommand{\xRightarrow}[2][]{\ext@arrow 0359\Rightarrowfill@{#1}{#2}}
\makeatother
\newcommand{\xdasharrow}[2][->]{
\tikz[baseline=-\the\dimexpr\fontdimen22\textfont2\relax]{
\node[anchor=south,font=\scriptsize, inner ysep=1.5pt,outer xsep=2.2pt](x){#2};
\draw[shorten <=3.4pt,shorten >=3.4pt,dashed,#1](x.south west)--(x.south east);
}
}
\subsection{The general translation}
In this section we bring together the modeling approaches described in the previous three sections and 
quickly summarize the translation from \dbnets to \nucpns with priorities. Specifically, we show that, given a \dbnet, 
it is possible to build a \nucpn that is weakly bisimilar to it.

Intuitively, $\cpn_\tau$ from Figure~\ref{fig:cpn-encoding} behaves just like the $\dbn_\tau$ in Figure~\ref{fig:dnet-generic} and hence LTSs of these 
two nets are bisimilar~\cite{Milner89}. Notice that, in order to correctly represent the behavior of $\dbn_\tau$, $\cpn_\tau$ includes many
intermediate steps that are, however, not relevant for comparing content of the states and behavior of the nets. For this we are going to 
resort to a form of bisimulation that allows to ``skip'' transitions irrelevant for the behavioral comparison~\cite{Milner89}. 
Specifically, given two transition systems 
$\tsys{}{1} = \tup{\states_1,s_{01},\trans_1,L}$ and $\tsys{}{2} = \tup{\states_2,s_{02},\trans_2,L}$ defined over a set of labels $L$, we call relation $wb\subseteq\states_1 \times \states_2$ a \emph{weak bisimulation} between $\tsys{}{1}$ and $\tsys{}{2}$ iff for every pair $\tup{p,q}\in wb$ and $a\in L\cup\set{\epsilon}$ the following holds:
\begin{inparaenum}[(1)]
\item if $p\xRightarrow{a}_1p'$, then there exists $q'\in\states_2$ such that $q \xdasharrow[->,>=latex]{~a~}_2 q'$ and $\tup{p',q'}\in wb$;
\item if $q\xRightarrow{a}_2q'$, then there exists $p'\in\states_1$ such that $p \xdasharrow[->,>=latex]{~a~}_1 p'$ and $\tup{p',q'}\in wb$.
\end{inparaenum}
Here, $\epsilon\neq a$ is a special silent label and $p \xdasharrow[->,>=latex]{~a~} q$ is a weak transition that is defined as follows:
\begin{inparaenum}[\it{i)}]
\item $p \xdasharrow[->,>=latex]{~a~} q$ iff $p (\xRightarrow{\epsilon})^* q_1\xRightarrow{a}q_2 (\xRightarrow{\epsilon})^* q$;
\item $p \xdasharrow[->,>=latex]{\ensuremath{~\epsilon~}} q$ iff $p (\xRightarrow{\epsilon})^* q$.
\end{inparaenum}
We use  $(\xRightarrow{\epsilon})^*$ to define the reflexive and transitive closure of $\xRightarrow{\epsilon}$.
We say that a state $p\in\states_1$ is weakly bisimilar to $q \in \states_2$, written $p \approx ^{wb} q$, if there exists 
a weak bisimulation $wb$ between $\tsys{}{1}$ and $\tsys{}{2}$ such that $\tup{p,q}\in wb$. 
Finally, $\tsys{}{1}$ is said to be weakly bisimilar to $\tsys{}{2}$ , written $\tsys{}{1}  \approx ^{wb} \tsys{}{2}$, if $s_{01}  \approx ^{wb} s_{02}$.
Let us now define a theorem that sets up the behavioral correspondence between \dbnets and \nucpns.

\begin{theorem}
Let $\dbn = \tup{\types,\pers,\dl,\cl}$ be a \dbnet with $\pers = \tup{\schema,\emptyset}$, $\dl=\tup{\queries^{UCQF^{\neq}},\actions}$ and $\cl = \tup{\places,\transitions,\inflow,\outflow,\rbflow,\coloring,\quass,\guass,\aass}$, and $\istate$ is the initial snapshot. 
Then, there exists a \nucpn $\cpn=\tup{\places\cup\places_{rel}\cup\places_{aux},\transitions\cup\transitions_{aux},\inflow,\outflow,\coloring,M_0}$ with
\begin{inparaenum}[(1)]
\item a set of relation places $\places_{rel}$ acquired from $\schema$,  
\item two sets $\places_{aux}$ and $\transitions_{aux}$ of auxiliary places and transitions (required by the encoding algorithm),
\end{inparaenum}
and such that $\tsys{\istate}{\dbn}\approx^{wb}_{flat}\tsys{M_0}{\cpn}$. 
\end{theorem}
The proof of the theorem is obtained inductively by modularly considering the encoding defined in Sections~\ref{sec:encoding-views}--\ref{sec:encoding-constraints}. Intuitively, the encoding lifts the persistence and data logic layers to the control layer, resulting in a ``pristine'' \nucpn. To show behavioral correspondence, one should make sure that states of $\tsys{\istate}{\dbn}$ and $\tsys{M_0}{\cpn}$ are comparable. 
This can be achieved by slightly modifying the notion of weak bisimulation in such a way that, for each $\tup{\tup{\I,m},M}\in wb$, we compare elements stored in $\I$ only with their ``control counterparts'' in $\places_{rel}$ of $M$, whereas $m\subseteq M$. 
Moreover, we assume that states of $\tsys{M_0}{\cpn}$ are restricted only to places in $\places\cup\places_{rel}$, that is, each marking $M$ shall reveal tokens stored only in $\places$ and $\places_{rel}$, and that when constructing $\tsys{M_0}{\cpn}$ all the auxiliary transitions of $\cpn$ (i.e., all the transitions within the grey lane in Figure~\ref{fig:cpn-encoding}) are going to be labeled with $\epsilon$.
Note that such an extended definition allows to establish equivalence not only in terms of behaviors of two systems, but also in terms of their (data) content.

\section{Conclusions}
We have shown that the large and relevant fragment of DB-nets employing unions of conjunctive queries with negative filters as database query language, can be faithfully encoded into a special class of Coloured Petri nets with transition priorities. Since the encoding is based on a constructive technique that can be readily implemented, the next step is to incorporate the encoding into the DB-net extension of CPN Tools \cite{RMRS18}, in turn making it possible to make the state-space construction mechanisms available in CPN Tools also applicable to DB-nets. It must be noted that, due to the presence of data ranging over infinite colour domains, the resulting state-space is infinite in general. However, in the case of state-bounded DB-nets \cite{MonR17}, that is, DB-nets for which each marking contains boundedly many tokens and boundedly many database tuples, a faithful abstract state space can be actually constructed using the same approach presented in \cite{CDMP17}. Interestingly, this can be readily implemented by replacing the ML code snippet dealing with fresh value injection with a slight variant that recycles, when possible, old data values that were mentioned in a previous marking but are currently not present anymore.

\bibliographystyle{splncs04}
\bibliography{mybib}
\end{document}